\documentclass[a4paper,11pt]{article}

\usepackage{amsthm,amsmath,amsfonts,amssymb}
\usepackage{bbm}
\usepackage{dsfont}
\usepackage{stmaryrd}
\usepackage{url}

\usepackage{epsfig}
\usepackage{figlatex}

\usepackage[ruled,vlined]{algorithm2e}
\usepackage{url}

\usepackage{multicol}

\usepackage{subfig}

\graphicspath{{Figures/}}

\numberwithin{equation}{section}
\theoremstyle{plain}
\newtheorem{theorem}{Theorem}[section]
\newtheorem{definition}[theorem]{Definition}
\newtheorem{remark}[theorem]{Remark}
\newtheorem{example}[theorem]{Example}
\newtheorem{proposition}[theorem]{Proposition}
\newtheorem{lemma}[theorem]{Lemma}
\newtheorem{corollary}[theorem]{Corollary}

\def\Blackboardfont{\mathbb}

\def\Space{\mathcal{S}}
\def\S{\mathcal{S}}

\def\Rb{\right\rrbracket}
\def\Lb{\left\llbracket}

\def\Evt{\mathcal{A}}
\def\newI{m'}
\def\newS{M'}
\newcommand{\argmin}{{\arg\min}}
\newcommand{\bydef}{\stackrel{\rm{def}}{=}}
\def\integer{{\Blackboardfont N}}
\def\real{{\Blackboardfont R}}
\def\R{{\Blackboardfont R}}
\def\PP{{\Blackboardfont P}}
\def\Z{{\Blackboardfont Z}}
\def\N{{\Blackboardfont N}}
\def\Q{{\Blackboardfont Q}}
\def\E{{\Blackboardfont E}}
\def\Act{{\mathit{Act}}}

\def\peq{\preceq}

\def\A{{\mathcal{A}}}

\def\CR{{\mathit{CR}}}
\def\edot{\boxdot}

\def\I{{\cal I}}

\newcommand{\ie}{{\it i.e.~}}
\newcommand{\iid}{{\em i.i.d.~}}
\newcommand{\abl}{\mathcal{B}}

\author{Ana {\sc Bu\v{s}i\'c}%
\thanks{INRIA/ENS, 23, avenue d'Italie, CS 81321, 75214 Paris Cedex 13, France. 
E-mail: {\tt Ana.Busic@inria.fr, pin@clipper.ens.fr}.}
\and
Bruno {\sc Gaujal}%
\thanks{INRIA and LIG, 
51, avenue Jean Kuntzmann
Montbonnont, France.
E-mail: {\tt bruno.gaujal@inria.fr}.}
\and
Furcy {\sc Pin}%
\footnotemark[1]
}

\begin{document}

\title{Perfect Sampling of Markov Chains with Piecewise Homogeneous Events} 

\date{}

\maketitle

\begin{abstract}
Perfect sampling is a 
technique that uses coupling arguments to provide a sample from the stationary 
distribution of a Markov chain in a finite time without ever computing the distribution.
This technique is very efficient if all the events in the system have monotonicity property.  
However, in the general (non-monotone) case, this technique needs to
consider the whole state space, which limits its application only to chains with a state space of
small cardinality. 
We propose here a new approach for the general case that only needs to consider two trajectories.  
Instead of the original chain, we use two bounding processes (envelopes) and we
show that, whenever they couple, one obtains a sample under the stationary distribution 
of the original chain.
We show that this new approach is particularly effective when the state space can be partitioned into pieces where envelopes can be easily computed.
We further show that most Markovian queueing networks have this property and we propose efficient algorithms for some of them.
\end{abstract}

\smallskip


\smallskip

{\noindent\bf AMS classification (2010):} Primary: 60J10, 65C05. Secondary: 68M20.
% 60J10    	Markov chains (discrete-time Markov processes on discrete state spaces)
% 65C05    	Monte Carlo methods
% 68M20    	Performance evaluation; queueing; scheduling

% \begin{table}[ht]
%   \centering
%   \caption{Notation Table}
%   \begin{tabular}{lrr}
%     \hline
%     State space    & $\S$ & \\
%     Number of dimensions & $d$ & \\
%     Dimension index & $i,j$ & \\
%     Other index & $k$ & \\
%     Set of points  & $U \subset \S$ & \\
%     Inferior bound & $m$ & \\
%     Superior bound & $M$ & \\
%     Interval & $[m,M]$ & (sometimes $[q,Q]$) \\ 
%     Event set & $\Evt$ & \\
%     Event      & $a \in \Evt$ & \\
%     Word & $a_{1 \rightarrow n}$ & \\
%     Word length & $n$ & \\
%     Transition: Action on points by events & $x \cdot a$ & \\
%     Transition: Action on sets by events & $U \cdot a$ & \\
%     Bounding transition: Action on intervals by events & $[m,M] \boxdot a$ & \\
%     Vector $(0,\dots,0,1 (i \text{th}),0,\dots,0)$
%     & $e_i$ &  \\
%   \end{tabular}
% \end{table}

%\tableofcontents

\section{Introduction}

Simulation approaches can be efficient ways  to
estimate the stationary  behavior of Markov chains  by providing
independent samples 
distributed according to their 
stationary distributions, even when it is
impossible to compute this distribution numerically.

Propp and Wilson used  a backward coupling scheme \cite{propp96exact}
to  derive a simulation algorithm - called PSA (Perfect Sampling
Algorithm)
in the following - providing  perfect samples (\ie whose
distribution is stationary) of the state of  a  {\it discrete time
Markov chain} 
over a finite state space $\Space$.
The main idea is to simulate trajectories  starting from  all states $x \in \Space$ at some time
in the past until time  $t=0$.  
If the state at time $0$ is the same for all trajectories, then the chain has {\it coalesced} 
and the final state has the stationary distribution of the Markov chain.
Otherwise, the simulations are started further in the past. 
Perfect  sampling procedures have been developed  in
various contexts (for more information, see the annotated bibliography~\cite{dbwilson}).

Although the Perfect Sampling Algorithm provides perfect samples for
all irreducible and aperiodic finite Markov chains, its complexity can be high. It
suffers from two drawbacks that jeopardize its 
applicability for very large chains.
\begin{itemize}
\item 
The first one is the fact that the coupling time can  be very large.
Some recent work  focused on the estimation  of the coupling time for
certain classes of Markov chains.
For example, it was  shown in \cite{Dopper} that Markov chains,
modeling a class of networks of queues with finite capacities, have a 
quadratic  coupling time with respect to capacities.

\item
The second factor in the complexity of PSA   is the fact that one
needs to run one simulation per  state in $\Space$, which is prohibitive for
most applications.  
Various techniques have been developed to
reduce the number of trajectories that need to be considered in the
coupling from the past procedure. A first crucial observation already
appears in \cite{propp96exact}: for a monotone Markov chain, one has to consider 
only extremal initial conditions.
For anti-monotone systems, an analogous technique that also considers 
only extremal initial conditions has been developed by 
Kendall \cite{Kendall-1998d} and H\"aggstr\"om and Nelander \cite{HaNe98}.

To cope with non-monotonicity, Kendall and M\o ller introduced 
in \cite{KendallMoeller-2000} the general idea of two bounding processes 
sandwiching all the normal trajectories, reducing the number of processes 
having to be computed to two. They applied the sandwiching technique to perfect sampling of point processes, 
together with the idea of a dominating chain that allowed to handle the infinite state space.
The idea of sandwiching processes was also applied  in communication networks \cite{BBT02, GaujalPerronninJDEDS}. 
%having to be computed to two, and applyied to Metropolis-Hastings simulations.
Huber \cite{Hu04} also introduced a similar idea of bounding chains for determining 
when coupling has occurred. In his model, the bounding chains are evolving in 
some bigger state space than the target Markov chain.
However, both  Kendall-M\o ller's and Huber's constructions are
model-dependent and do not have an algorithmic counterpart so that they   are not straightforward to apply in general.

\end{itemize}

In this paper we introduce a new perfect sampling algorithm 
(EPSA: Envelope Perfect Sampling Algorithm) that uses the 
lattice structure of the state space to design an effective and automatic construction 
of a bounding interval chain that can be expressed using only two trajectories. 
This is done by computing from the past, a lower and an upper bound of all
trajectories of the Markov chain. 
Whenever the two bounds meet, the coupling state is distributed 
according to the stationary distribution of the chain.
The idea of bounding interval chain is close to sandwiching
processes in Kendall and M\o ller \cite{KendallMoeller-2000}.
However, their approach was not constructive and the sandwiching processes were computed ad-hoc for each application.

Our aim is to propose, under a constructive approach, several algorithms 
to compute bounding interval chains that are easily computable and as tight as possible.
% This approach is closest to Kendall and M\o ller \cite{KendallMoeller-2000}.
% 
% Our aim is to propose, under a constructive approach, several algorithms 
% to compute bounding interval chains that are easily computable and as tight as possible.
%
%approche algorithmique constructive
%
%
This is done in two cases.
The first (easy) case is when events are {\it almost state space
homogeneous} (defined in Section \ref{sec-ashes}). 
In that case, we provide an algorithm that computes 
the update rule of the bounding interval chain. % in linear time.
The second case is more difficult since events are assumed to be 
piecewise defined (Section \ref{sec:partition}) over arbitrary polytopes (Section \ref{sec:lp}).
In that case we show how to obtain an  interval containing
all trajectories by solving a set of linear programs.
Since the worst case complexity of linear programs may be too high, 
we show how to avoid solving these, at the expense of obtaining 
bounding intervals that may be less tight.

In the last part of the paper (Section \ref{sec-ex}),  
we consider applications to queueing networks. 
We exhibit several events (such as negative customers, fork and join,  
batch routing) that satisfy almost state space homogeneity. 
We also provide numerical evidence of the very good performance of EPSA
in these cases, by computing the coupling time experimentally.

We also show examples (such as generalized join the shortest queue)
that are  piecewise homogeneous, % on polytopes, 
where the smallest interval containing all trajectories can also
be computed in linear time in the number of queues.
Our numerical experiments show that EPSA behaves just as well as in
the easy case of almost space homogeneous events.

\section{Definitions and Basic Properties}
\label{sec-def}

If not stated otherwise, all real (or integer) vectors are row vectors
and $v^t$ denotes the transposed vector. 
For $x, y\in \real^d$, we denote  $x \lor y \bydef
(\max\{x_1,\;y_1\}, \ldots, \linebreak \max\{x_d,\;y_d\})$,  
and $x \land y \bydef (\min\{x_1,\;y_1\}, \ldots, \min\{x_d,\;y_d\})$.
To avoid multiple brackets, for $x, y,z\in \real^d$, we will write: 
$$x \lor y \land z \bydef (x \lor y) \land z.$$
Note that $(x \lor y) \land z \not= x \lor (y \land z)$;
the relations involving operators $\lor$ and $\land$ will be evaluated from left to right.
For any $x \in \real^n$, $||x||_1 = \sum_{i=1}^n |x_i|$,
and for any set $U$, $|U|$ denotes its cardinality.  

Throughout the paper, we consider discrete time homogeneous Markov
chains (DTMC) with a finite state space $\Space$.
Since we only consider  finite state spaces, the continuous time models
can be easily handled using the same techniques, after uniformization. 

\subsection{Markov Automaton}

The evolution of a finite DTMC can always be obtained using a finite number of discrete 
events (or actions). To formalize this, we introduce the notion of Markov automaton. 
Markov automatons are similar  to Generalized Semi Markov Processes \cite{cassandras} 
with a focus on state changes rather than on time.

\begin{definition}
  A {\em Markov automaton} is a tuple $\A=(\Space,A,D,\cdot)$ where $\Space$ is
  a finite state space, $A$ is an alphabet called the {\em set of events}, 
  $D$ is a probability distribution on $A$,
  and $\cdot$ is a right action by the letters of $A$ on the 
states in $\Space$ and is called the {\em transition function} :
  \begin{equation*}
    \cdot \quad : \quad  
    \Space \times A \rightarrow \Space , (x,a) \mapsto x \cdot a .
  \end{equation*}
  Equivalently, this action can be most naturally extended to 
  a right action by words $u_{1 \rightarrow n} \bydef u_1 \dots u_n$ in $A^n$ on $\Space$ 
  for any $n \in \N$:
  \begin{equation*}
    \cdot \quad : \quad  
    \Space \times A^n \rightarrow \Space ,
    (x,u_{1 \rightarrow n}) \mapsto x \cdot u_{1 \rightarrow n} \bydef x 
    \cdot u_1 \cdot u_2 \cdot \ldots \cdot u_n .
  \end{equation*}
\end{definition}

A Markov automaton $\A=(\Space,A,D,\cdot)$ naturally induces 
a Markov chain with the following construction:

Let $(u_1,\dots,u_n,\dots)$ be an infinite sequence of random letter of $A$ \iid
distributed according to $D$. 
We consider this sequence as an infinite word $u \bydef u_1 \dots u_n \dots$ on $A$.
Then for any $x_0 \in \Space$, the random process 
$(X_n \bydef x_0 \cdot u_{1 \rightarrow n})_{n \in \N}$
is a Markov chain issued from $x_0$ with probability transition matrix $P$ given by
\begin{equation}\label{eq:markov_automaton}
  \text{for all } x,y \text{ in } \Space, ~~ P(x,y) = 
\underset{x \cdot a = y}{\sum_{a \in A}} \PP_D(a) .
\end{equation}
We will say that the Markov chain $(X_n)$ is \textit{generated} by $\A$ and $u$.
Moreover, with this construction, we can build a family of Markov chains 
$\{(X_n(x) = x \cdot u_{1 \rightarrow n})_{n \ge 0} \mid x \in \Space\}$ 
starting from each state $x \in \Space$. 
We will refer to this family of Markov chain as the {\em grand coupling} 
generated by $\A$ and $u$.
We will say that the grand coupling has {\em coalesced} at time $n$
if all the Markov chains of the family has reached the same state,
\ie when $\Space \cdot u_{1 \rightarrow n}$ is reduced to a singleton.

Conversely, for any probability transition matrix $P$ on a finite space state $\S$,
it is easy to see that there exists a Markov automaton $\A=(\Space,A,D,\cdot)$ such that 
\eqref{eq:markov_automaton} holds, \ie such that $\A$ generates a Markov chain
on $S$ with same transition matrix $P$, but that automaton is in general not unique.
However as we will see in our examples, such representation of Markovian systems
with Markov automatons naturally arises for many systems such as queueing networks.

\subsection{Perfect Sampling} 

Let $\left\{X_n\right\}_{n \in \integer}$ be an irreducible
and aperiodic DTMC with a finite state space
$\Space$ and a  transition matrix $P$.
Let  $\pi$ denote the steady state distribution of the chain: $\pi = \pi P$.

The Perfect Sampling Algorithm (PSA) was first used by Propp and Wilson.
It allows one to draw in finite time steady state distributed variable using coupling from the past.

Consider a Markov automaton $\A=(\Space,A,D,\cdot)$ and
$(u_0,u_{-1},\dots,u_{-n},\dots)$ a sequence of events \iid with distribution $D$ on $A$.

\begin{theorem}[Propp and Wilson \cite{propp96exact}] 
\label{thm:psa}
There exists  $\ell \in \N$ such that 
$$\lim_{n\to \infty} \big| \Space \cdot u_{-n+1\to 0}
\big| =\ell \; \textrm{almost surely}.$$

The grand coupling generated by $\A$ and $u$ couples if $\ell = 1$. 
In that case, let 
$\tau \bydef \inf \bigg\{ n \; : \; \big| \Space \cdot u_{-n+1\to0} \big| = 1 \bigg\}$
be the coupling time of the chain. Then $\E(\tau) < \infty $ and 
$\Space \cdot u_{-\tau+1 \rightarrow 0}$ is steady state distributed. 
%Furthermore, given  an irreducible  transition matrix $P = (p_{i,j})$, it is 
%possible to construct a transition function $\Phi$ that couples so
%that the Perfect Sampling Algorithm 1 (PSA)  can always be constructed.
\end{theorem}

\begin{algorithm}
\KwData{
I.i.d. events  $\left\{u_{-n}\right\}_{n\in \integer} \in \Evt^\N$
}
\KwResult{A state $x^* \in \Space$ generated according to the stationary
 distribution of the Markov chain}
\Begin{
$n:=0$\;
\lForEach{state $x \in \Space$} {$S[x] := x$\;}
\tcp{$S[x]$ is the state of the trajectory issued from $x$ at time $-n$.}
\Repeat{$\left|\left\{S[x], x \in \Space \right\}\right|=1$  \newline
($x^*$ denotes this unique value)}
{ 
%$y(\cdot) := y(\Phi(\cdot,e_{-n}))$;\\
\lForEach{state $x \in \Space $} {$R[x] := S[x \cdot u_{-n}]$;}
$S := R$\;
$n:=n+1$;}
\Return{$x^*$}}
\caption{Perfect Sampling Algorithm (PSA)}
\label{psa1}
\end{algorithm}

The algorithmic counterpart (PSA) of Theorem \ref{thm:psa} is given in 
Algorithm~\ref{psa1}. The main drawback of PSA algorithm is the fact
that one needs to simulate one Markov chain starting from each state in $\Space$, 
that could be too large for a practical use of the algorithm.

Several approaches have been used to overcome this problem.
The main one for a 
partially ordered state space $(\Space, \preceq)$ and monotone events
is already present in \cite{propp96exact}.  
\begin{definition}
An event $a \in \Evt$ is said to be monotone if $x \preceq y \;
\Longrightarrow x \cdot a \preceq y \cdot a$. 
An event $a \in \Evt$ is said to be anti-monotone if 
$x \preceq y \; \Longrightarrow x \cdot a \succeq y \cdot a$.
\end{definition}
If all events are monotone, then it is possible to 
generate a steady state by starting  Algorithm~\ref{psa1} with maximal and
minimal states only \cite{propp96exact}. 
This technique has been successfully used in \cite{V_psmsfr_05} to
construct PSA for 
network of queues. Similarly, an efficient perfect sampling
algorithm for the chains with anti-monotone events was given in \cite{HaNe98}.
 
When events are neither monotone nor anti-monotone, one can still use
monotone bounds, as 
in \cite{KFPVV_psmsb_07}, but this technique provides a sample from a
bounding chain and not the initial one.   
We give in this paper an algorithm for perfect sampling of general (non-monotone) chains.

\section{Envelopes}\label{sec:env}
The approach proposed here generalizes what has been done in
\cite{GP_cfphmf_07} and \cite{BGV08} to sample non-monotone Markov chains.
It uses a bounding chain method similar to the one employed by Kendall and M{\o}ller in \cite{KendallMoeller-2000}. 
The sandwiching chains in \cite{KendallMoeller-2000} are in general difficult to compute, 
except for some models for which effective ad-hoc rules can be derived.
Our approach is more algorithmic. We introduce  a general framework 
that provides efficient algorithms for a wide variety of events,
including many queueing events.

\subsection{Bounding Interval Chains}

We assume from now on that the state space $\Space$ is equipped with a 
lattice order relation $\peq$. 

For $m, M \in \Space$, we will denote by 
$[m, M] \bydef \{x \in \Space \; : \; m \peq x  \peq  M\}$
the \textit{lattice interval} (or simply \textit{interval}) between the endpoints $m$ and $M$
(note that $[m,M] \neq \emptyset$ if and only if $m \preceq M$).
We introduce $\I$ as the set of all nonempty lattice intervals:
$\I = \left \{[m, M] \; : \; m, M \in \Space, \; m \peq  M \right \}$.

We define bounding interval chains as follows.

\begin{definition}
  Given a grand coupling $\{ (X_n(x))_{n \ge 0} \mid x \in \Space \}$, 
  we call {\em bounding interval chain of that grand coupling}  
  any Markov chain of nonempty intervals $([m_n,M_n])_{n \ge 0}$
  such that for all $x$ in $\Space$ and all 
  $n \ge 0$, $X_n(x) \in [m_n,M_n]$.
\end{definition}

In particular we notice that when $m_n=M_n$, the grand coupling inside has necessary coalesced.
Thus, by constructing efficiently such bounding intervals,  
we can detect easily if a grand coupling has coalesced and apply it to PSA.

We now construct the tightest bounding interval chain computable step by step 
for a given grand coupling, that  we will call envelope chain.

For any set $U \subset \Space$, we denote by $\Lb U \Rb$ and we call {\em envelope of $U$}
the smallest interval of $\Space$ (for inclusion) containing $U$ :
$ \Lb U \Rb \bydef  [\inf(U) , \sup(U)] $.
This is illustrated by Figure \ref{fig:interval}.
We introduce a new transition operator $\edot : \I \times \Evt \to \I$
called \textit{envelope transition} 
that transforms intervals into intervals 
and is defined by : for all $[m, M] \in \I$ and $a \in \Evt$:
\begin{equation}
  \label{eq:env}
        [m, M] \edot a  \bydef \Lb [m,M]\cdot a \Rb =  \left [ \inf_{m \peq x\peq
            M } \left \{x \cdot a\right \}, \;   
          \sup_{m\peq x\peq M} \left \{x \cdot a\right \} \right ].
\end{equation}
The construction of $\edot$ is illustrated on a simple example
in Figure \ref{fig:envelope}.

\begin{figure}[t]
\begin{multicols}{2}
  \includegraphics[width=.30\textwidth]{interval.fig}
  \caption{Intervals in dimension two for the natural product order. 
    The set $U$ is the set of all the black
    points. The interval  $\Lb U \Rb$ is displayed as a rectangle.}
  \label{fig:interval}
  \includegraphics[width=.27\textwidth]{enveloppe.fig}
  \caption{Envelopes of the action of  $a$ on an interval. The
    interval $[m,M]$ is displayed as  black points. The action of $a$   is
    displayed as arrows. The envelope interval $[m',M']=[m,M]\edot 
    a$ is displayed as a rectangle.}
  \label{fig:envelope}
\end{multicols}
\end{figure}

As with `$\cdot$`, we extend the operator to a finite word of events
$u_{1\to n} = u_1 \dots u_n\in \Evt^n$: 
$$[m, M] \edot u_{1\to n} \bydef [m, M] \edot u_{1}  \edot u_{2} \edot \dots \edot u_{n}.$$ 
Let us call $\bot \bydef \inf \Space$ (resp . $\top \bydef  \sup
\Space$)  the bottom  
(resp.  top) element of $\Space$.
The process $$[m_n,M_n] \bydef [\bot,\top] \edot u_{1 \to n}$$ is a Markov chain
over the state space   $\Space \times \Space$, called the {\em envelope chain},
and is a bounding interval chain of the grand coupling 
$\{(x \cdot u_{1 \rightarrow n}) \mid x \in \Space\}$.

It is also the tightest bounding chain computable step by step, in the sense
that if, at time $n$, we only know that $S \cdot u_{1 \rightarrow n}$ 
is in $[m_n,M_n]$, then the tightest interval containing $S \cdot u_{1 \rightarrow n+1}$ 
we can compute at time $n+1$ is $[m_n,M_n] \edot u_{n+1}$.
However, it is not tight in the stronger sense of
$[m_n,M_n]$ being equal to $\Lb S \cdot u_{1 \rightarrow n} \Rb$, 
that does not hold in general.

Notice that  the lower envelope $(m_n)_{n\in\integer}$ alone is not a
Markov chain, neither is the upper one $(M_n)_{n\in\integer}$, since 
they depend on each other.

We can then use the envelope process to detect coupling in PSA.

\begin{theorem}\label{th:env}
Assume that  $[m_n,M_n]$  hits the set of single point intervals
$\mathcal{X} = \left \{[x,x] \; : \; x \in \Space \right \}$ 
in finite time:
\begin{eqnarray*}
{\tau_e} &\bydef& \min \bigg\{ n \; : \; [\bot,\top] \edot u_{-n+1\to0} \in \mathcal{X}\bigg\}, 
\end{eqnarray*}
then $\tau_e$  is a  backward coupling time of the envelope
chain. The state defined by  $[\bot,\top] \edot u_{-\tau_e+1\to 0}$
has the  steady state distribution of DTMC $\{X_n\}_{n\in \integer}$.
\end{theorem}
\begin{proof}
The proof simply uses the fact that the set $\Space \cdot u_{-n+1\to 0}$ 
is included in $[\bot,\top] \edot u_{-n+1\to 0}$.  
Consequently, if the latter  is reduced to one point, 
so is the former and the grand coupling has coalesced.
\end{proof}

Algorithm \ref{psa1}  can be adapted to the envelope simulation:
start the simulation with only two states, $\bot$ and $\top$, and iterate
using $\edot$ and an \iid sequence of events.
This new algorithm, called EPSA (Envelope Perfect Sampling Algorithm),
is given in Algorithm~\ref{al:EPSA}. 
The reason why we double $n$ at each loop of the algorithm is that
we need to compute $S \edot u_{-n+1 \rightarrow 0}$ in each loop,
which corresponds to $n$ iterations of $\edot$.
While increasing $n$ by $1$ at each loop would lead to a quadratic cost
in $n$, doubling it keeps the complexity linear.

 \begin{algorithm}
\KwData{
I.i.d. events  $\left\{u_{-n}\right\}_{n\in \integer} \in \Evt^\N$; 
the (pre-computed) envelope operation $\edot$ 
}
\KwResult{A state $x^* \in \Space$ generated according to the stationary
 distribution of the Markov chain}
  \Begin{
$n=1$;  $m :=   \bot$; $M :=  \top$\;
  \Repeat{$m = M$}
{ \For{$i=n-1$ {\em  \textbf{downto}} $0$}
{ $[m, M] :=[m, M] \edot u_{-i}$ \;}
$n:=2n$\;}
$x^* := m$\;
 \Return{$x^*$\;}}
\caption{Envelope Perfect Sampling Algorithm (EPSA)}
\label{al:EPSA}
\end{algorithm}
 
EPSA is a generalization of both monotone and anti-monotone  
sampling algorithm~(see \cite{propp96exact,HaNe98}): 
\begin{itemize} 
 \item Monotone events: if an event $a \in \Evt$ is monotone, 
then for any $m\peq M$, 
\begin{equation}
\label{eq:mono}
[m, \; M] \edot a  = [m \cdot a, \; M \cdot a].
\end{equation}
 \item Anti-monotone events: if an event $a \in \Evt$ is anti-monotone, 
then for any $m\peq M$, $[m, \; M] \edot a  = [M \cdot a, \; m \cdot
a]$.  
\end{itemize}

In the next section we discuss the complexity of EPSA, 
that  depends on the coupling time of the envelope chain, but also
on the complexity of the computation of the envelope transition.
The efficiency of the (anti-) monotone case is due to the fact that  
when $`\cdot`$ is (anti-)monotone, the complexity of $\edot$ is 
the same as using $`\cdot`$ twice.

Note however that the construction of the envelopes depends 
on the representation of the Markov chain by a Markov automaton 
that  is not unique. 
Different automatons representing the same chain 
would lead to different envelope chains with different  
coupling properties (one may coalesce almost surely and the other not, or if
they both coalesce their coupling times may be different). 
The complexity of the envelope transition may also differ depending on the representation chosen.

\subsection{Complexity of EPSA}
\label{ss:CI}

The envelope approach may not gain over the 
general PSA because of two drawbacks:

\begin{itemize}
\item[$(D_1)$] The assumption that $[m_n, M_n]$ hits the set of 
  single point intervals $\mathcal{X}$
  in Theorem~\ref{th:env} may not hold.
\item[$(D_2)$] Even if the assumptions for Theorem~\ref{th:env} are
  satisfied, the complexity of the algorithm may be prohibitive. 
The average complexity of EPSA is $O({\cal C}_e \times \E(\tau_e)  )$
where $\tau_e$ is the 
number of iterations of the main loop of EPSA (called the coupling
time in the following) and  
${\cal C}_e$ is the complexity (number of elementary operations) of
computing $[m,M]\edot a$.
Meanwhile, the average complexity of the classical PSA is 
$O( {\cal C} \times |\Space | \times \E(\tau) )$, 
where $\tau$ is the coupling time of PSA, $|\Space |$
the cardinality of the state space and  ${\cal C}$ the complexity of the
computation of $x\cdot a$.
\end{itemize}

\begin{remark}
It is important to notice that the operation $\edot$ can be replaced by any 
over-approximation $\odot$ such that for all interval 
$[m,M]$, $[m,M]\edot a \subseteq [m,M]\odot a$,
without altering the correctness of the algorithm.
By doing so, we would still obtain a bounding interval chain,
however its coupling time might be longer than for the envelope chain, or the
bounding interval chain might not couple at all. 
There is then an obvious tradeoff between 
computing simple large bounds quickly with an increased coupling time,
or computing tight bounds slowly with a potentially shorter coupling time.
\end{remark}

\begin{remark}
Actually, even if the envelope chain does not couple, it is still 
possible to obtain perfect samples in finite time by using the 
splitting algorithm \cite{BGV08}. 
The splitting algorithm is hybrid: it first runs EPSA when 
the envelopes are too far apart and switches to the usual PSA algorithm 
as soon as the number of states inside the envelopes becomes manageable.
\end{remark}

The rest of this paper focuses on improving the cost ${\cal C}_e$ of computing envelopes
under several assumptions except for Section \ref{sec:expansive} and 
numerical examples in Section \ref{sec-ex} that discuss the coupling time issue.

Indeed, if ${\cal C}_e$ is not too large compared to ${\cal C}$ 
(as it is the case for the various transitions proposed in this article),
then the comparison of the two methods is essentially a comparison
between the coupling time of EPSA ($\E (\tau_e)$) 
and the coupling time of 
PSA multiplied by the cardinality of the state space ($|\Space |
\times \E (\tau)$).

\section{Almost Space Homogeneous Events (ASHEs)}
\label{sec-ashes}

We now present a family of simple events called
\textit{Almost Space Homogeneous Events (ASHEs)} for which the corresponding
envelope transitions are easy to compute. These events will be the first
brick to build more elaborate ones.
Intuitively, these events consist in moving homogeneously every point
of the space in the same direction, with some simple conditions 
on the boundaries of the state space. 

\subsection{Computing the envelopes for ASHEs}
\label{compute-ashes}

We assume from now on that $\Space$ is an interval of $\Z^d$:
$\Space = [0,C]$, where $C = (C_1,\ldots,C_d) \in \Z^d$ 
endowed by the usual product partial order: 
$ x \leq y  \; \Leftrightarrow \; x_i \leq y_i, \; \forall i $. 
Therefore for any $m,M \in \Space$, we have $[m,M]=[m_1,M_1] \times \dots \times [m_d,M_d]$.

\begin{definition}
 An event $a$ is an {\it almost space homogeneous event} (ASHE) 
if there exists a vector $v \in \Z^n$ and a binary relation $\cal{R}$
(called {\em blocking relation}) on the set of components $\{1,\dots,d\}$ 
such that for all $x$ in $\Space$ we have the following property:

\begin{itemize}
  \item[$(\mathcal{P})$]
  Let $CR(x) \bydef \{i ~:~ x_i+v_i \not\in [0,C_i] \}$ be the set of the critical
  components (\ie where $x+v$ is out of the state space), 
  and $B(x) \bydef  \{ i ~:~ \exists j \in CR(x),~ (j,i) \in \mathcal{R} \}$
  the set of 
  all blocked components in state $x$.
  Then for all $i$ :
  \begin{equation}\label{eq:evt}
    (x \cdot a)_i = 
    \begin{cases}
      x_i, & i \in B(x), \\
      (x_i + v_i) \lor {0} \land {C_i}, & i \not\in B(x).
    \end{cases}    
  \end{equation}
\end{itemize}
In particular, when $x+v \in S$, we have $x \cdot a = x + v$.

We call $v$ the {\em direction vector} of event $a$. 
\end{definition}

\begin{remark}
  The blocking relation $\cal{R}$ needs only to be specified on 
  $Act \bydef \{i \mid v_i \neq 0 \}$ the set of {\em active components}
  of the transition, since the event has an effect only along these components.
\end{remark}

The cases $B(x) = \emptyset $ or $B(x) = d$  are  interesting
particular cases that correspond respectively to the  case where,  
if $x+v \not\in \Space$, then the event %touches 
is truncated at
the boundary of the  
state space: $x \cdot a = (x + v) \lor 0 \land C$, 
or the event is blocked: $x \cdot a = x$.

\begin{example}
\label{ex:fork}
Here is a more complex case in the context of queues.
Let us consider a Markovian queueing network made of 3 finite queues
and  exponential service times and  a fork node: one departing  packet from
queue 1 is duplicated 
into two packets joining queues 2 and 3.
This event is an ASHE, corresponds to adding a vector $v =  (-1, +1,
+1)$ to all 
states except on the boundaries.
The critical cases are when the first queue is empty or  when one of
the next queues is full.
If the first queue is empty, then the event is blocked  so that
$(1,2)$ and 
$ (1, 3)$ are in ${\cal R}$.
If either queue 2 or queue 3 are full, then the packet actually
leaves queue 1 and does not join the other queues (both duplicates
are lost).
This means that both $ (2,3), (3,2)$ are in ${\cal R}$.
These are the only constraints, thus ${\cal R} = \{ (1,2), (1,3), (2,3),
(3,2) \}$.
Other cases could also be taken into account, for example,
the case where a  full queue does  not prevent the arrival  in the other
queue, corresponding to ${\cal R} = \{ (1,2), (1,3)\}$,
or  when a full queue blocks the departure from the first queue,
corresponding to ${\cal R} = \{ (1,2), (1,3), (2,1), (2,3),
(3,1), (3,2) \}$.
\end{example}

\begin{figure}[hbtp]
  \centering
  \includegraphics[width=0.5\textwidth]{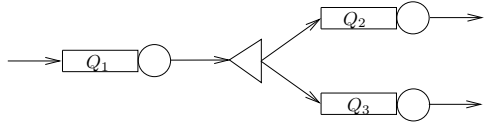}
  \caption{Fork node}
  \label{fig:fork}
\end{figure}

The computation of envelopes for a fixed ASHE $a$ is given in Algorithm~\ref{al:ashes}.

\begin{algorithm}[htbp]
\KwData{
States $m$ and $M$ (current infimum and supremum state)\;
An ASHE $a$ with direction vector $v$ and blocking relation $\mathcal{R}$\;
}
\KwResult{$[\newI, \newS] = [m, M] \edot a$ (lower and upper envelope).}
\Begin{
$\Act = \{i \; : \; v_i \not= 0\}$\;
$\CR(m) = \{ i \in \Act\; : \; m_i + v_i \not\in [0,C_i]\}$\;
$\CR(M) = \{ i \in \Act\; : \; M_i + v_i \not\in [0,C_i]\}$\;
$X = \{j \in \Act \; : \; \exists i \in \CR(m) \cup \CR(M)$ 
such that $(i, j) \in {\cal R}\}$\;
$Y = \{j \in \Act \; : \; \exists i \in \CR(m) \cap \CR(M)$ 
such that $(i, j) \in {\cal R}\}$\;
\tcp{Note that $Y \subset X$}
\For{$j\in \Act\backslash X$}{
\tcp{This is the case $j\not\in B(m) \cup B(M)$}
$\newI_j = (m_j + v_j) \lor 0 \land {C_j}$;  
$\newS_j = (M_j + v_j) \lor 0 \land {C_j}$\;
}
\For{$j\in Y$}{
$\newI_j=m_j$; $\newS_j= M_j$\;
}
\For{$j\in X\backslash Y$}{
\If{
   $\exists i, \; i \not=j$ such that  $(i, j) \in {\cal R}$ and $i
   \in \CR(m) \cup \CR(M)$}{  
  \tcp{component $j$ is blocked by a component $i$ different from $j$} 
 
\If{$v_j < 0$}{ 
$\newI_j = (m_j + v_j)\lor {0}$; $\newS_j =M_j$\;
}
\Else{
$\newI_j = m_j$; $\newS_j =(M_j + v_j) \land {C_j}$\;

}
 }
 \Else{
\tcp{only $j$ is blocking $j$}
 \If{$v_j < 0$}{
 $\newI_j =0$;  $\newS_j = (M_j+v_j) \lor (-v_j - 1)$\;
 }
 \Else{
 $\newI_j = (m_j+v_j) \land (C_j-v_j+1)$; $\newS_j =C_j$\;
 }
 }
}
\For{$j\not\in \Act$}{
$\newI_j=m_j$; $\newS_j= M_j$\;
}
}

\caption{Envelope computation for an ASHE $a \in \Evt$}
\label{al:ashes}
\end{algorithm}

\begin{theorem}
 \label{thm:ashes}
Algorithm \ref{al:ashes} computes 
$[\newI, \newS] = [m, M] \edot a$ in $O(d^2)$ elementary operations.
\end{theorem}
The proof is given in Appendix \ref{app:ashe}.

\subsection{Nonexpansive properties of ASHEs}
\label{sec:expansive}

In the previous Section, %\ref{compute-ashes}
we have shown that computing the envelopes 
for ASHEs can be done efficiently, so that the term ${\cal C}_e$ in the complexity is in $O(d^2)$. 
This section gives some insights on the coupling time.
Since our approach is general, no tight bound on the coupling  can be derived. 
The most immediate property is that, as mentioned in  Section \ref{sec:env}, 
the coupling time of envelopes is larger than the coupling time of the original chain.
However, as seen in the experimental section \ref{sec-ex}, the difference between two 
coupling times is not large.
This can be explained in part by several nonexpansive properties of ASHEs.
In this section we show that the size of the envelope
interval cannot increase more than linearly in the worse case, and
does not increase (or  even contracts) under  some conditions on the blocking relation.
In the latter case, the contracting property also gives some insight of the
splitting condition.

Assume that $a \in \Evt$ is an ASHE given by \eqref{eq:evt}.
We first give a useful lemma about the set of blocked components of an ASHE.

\begin{lemma}
  \label{lem:subset}
  Let $U$ be a subset of $\Space$ and let $ [m,M]= \Lb U \Rb$. We have 
  $$\bigcup_{x \in U}B(x) = B(m)\cup B(M).$$
\end{lemma}

\begin{proof}
We first prove the inclusion $\subseteq$ :
Let $x \in U$ and $j \in B$. Then there is $i \in CR(x)$ such that $(i,j) \in {\cal R}$.
Now $i \in CR(x)$ implies that $i \in CR(m) \cup \CR(M)$ 
(as $m_i + v_i \leq x_i + v_i \leq M_i +v_i$), thus $j \in B(m) \cup B(M)$. 
We now prove the other inclusion. Let $j \in B(m)$, there is a $i$ such that
$(i,j) \in \mathcal{R}$ and $m_i + v_i \notin [0,C_i]$. Since $[m,M]=\Lb U \Rb$,
there exist $x \in U$ such that $x_i=m_i$, and thus $x_i+v_i \notin [0,C_i]$
and $j \in B(x)$. This shows $B(m) \subseteq \bigcup_{x \in U}B(x)$, and the same
reasoning shows $B(M) \subseteq \bigcup_{x \in U}B(x)$.
\end{proof}

The following proposition shows the nonexpansive property in the case
without blocking. 
For $m, M \in \Space$ such that $m\leq M$, denote by $[\newI, \newS] =
[m, M] \edot a$.    
\begin{proposition}
\label{prop:part}
If for $m, M \in \Space$ such that $m \leq M$,  $B(m) \cup B(M) =
\emptyset$, then:  
 $$||\newS - \newI||_1 \leq ||M-m||_1.$$
\end{proposition}

\proof
If $B(m) \cup B(M) = \emptyset$, then by Lemma \ref{lem:subset}, 
$\cup_{x \in [m,M]} B(x) = B(m) \cup B(M) = \emptyset$. Then
(\ref{eq:evt}) becomes 
$x \cdot a = (x + v) \lor {0} \land {C}, \; \forall m \leq x \leq M$, 
that  is a monotone function of $x$ (\ie event $a$ is monotone). Therefore,  
$||\newS - \newI||_1 = ||M \cdot a - m \cdot a||_1  \leq ||M-m||_1.$
\endproof

\begin{corollary}
\label{cor:part}
If ${\cal R} = \emptyset$, then for any $m, M \in \Space$ such that $m \leq M$: 
$$||\newS - \newI||_1 \leq ||M-m||_1.$$
\end{corollary}

The next proposition gives an upper bound %for norm-$1$ after one iteration 
in the case where each component can only be blocked by itself.
In that case, relation ${\cal R}$ 
is defined by the set $\abl$ of all auto-blocking components: 
${\cal R} = \{(i,i) \; : \;  i \in \abl\}$.
\begin{proposition}
\label{prop:rej}
If ${\cal R} = \{(i,i) \; : \;  i \in \abl\}$, 
for some subset $\abl \subset \{1, \ldots, d\}$, then:
\begin{eqnarray*}
||\newS - \newI||_1 & \leq & 
\sum_{i \not\in \abl} (M_i - m_i) + \sum_{i \in \abl} \max\{M_i-m_i, |v_i|-1 \}\\
& = & ||M-m||_1 + \sum_{i \in \abl} \left[|v_i| - M_i + m_i - 1\right]\lor 0.  
\end{eqnarray*}
\end{proposition}
\proof
Since there are no interactions between components, we can consider
them separately.  
Thus, for $i \not\in \abl$:
$\newS_i - \newI_i \leq M_i-m_i$, as in Proposition~\ref{prop:part}.
For $i \in \abl$, we have three different cases:
\begin{itemize}
 \item For all  $x \in [m, M]$, $x_i + v_i \in [0, C_i]$. This is
   equivalent to $i \not\in CR(m) \cup CR(M)$.   
Then $(x \cdot a)_i = x_i + v_i, \; \forall x \in [m, M]$
and $\newS_i - \newI_i = M_i-m_i.$
\item For all $x \in [m, M]$, $x_i + v_i \not\in [0, C_i]$. This is
  equivalent to $i \in CR(m) \cap CR(M)$.   
Thus $(x \cdot a)_i = x_i, \; \forall x \in [m, M]$
and $\newS_i - \newI_i = M_i-m_i.$ 
\item There are $x, y \in [m, M]$ such that $x_i + v_i \in [0, C_i]$
and $y_i + v_i \not \in [0, C_i]$. Suppose that $v_i > 0$ (the  case
$v_i < 0$ is symmetrical). Then  
 $i \in CR(M) \backslash CR(m)$.  
Denote by $x'_i = C_i - v_i$. Then $m_i \leq x'_i <  M_i.$
We have:
$$\newI_i = (m_i+v_i) \land  (x'_i+1), \quad \newS_i = C_i.$$
There are two cases:
\begin{itemize}
\item $m_i+v_i \leq x'_i+1$.
Then $\newS_i - \newI_i = C_i - (m_i+v_i)
= x'_i - m_i \leq M_i - m_i-1.$
\item $m_i+v_i > x'_i+1$. Then $\newS_i - \newI_i = C_i -(x'_i+1) = v_i - 1.$
\end{itemize}
\end{itemize}
Thus $\newS_i - \newI_i \leq \max\{M_i - m_i, |v_i|-1\}.$
\endproof
The above result tells us that an event without blocking interactions
between different components  
(\ie $(i, j) \in {\cal R} \Longrightarrow i=j$)
is nonexpansive in components that are not auto-blocking. For
components that are auto-blocking 
such event is nonexpansive  when $M_i - m_i \geq |v_i|-1$.

\begin{corollary}
If ${\cal R} = \{(i,i) \; : \;  i \in \abl\}$, for some subset $\abl \subset \{1, \ldots, d\}$, 
and $|v_i| = 1, \forall \; i \in \abl$, then
$||\newS - \newI||_1 \leq ||M-m||_1, \;  m \leq M$. 
\end{corollary}

We give now an upper valid bound for any ASHE. 
\begin{proposition}
\label{prop:block}
For any ASHE '$a$' given by (\ref{eq:evt}), we have:
$$||\newS - \newI||_1 \leq 
||M-m||_1 + ||v||_1 - 1.$$
\end{proposition}
\begin{proof}
For $i \in \{1, \ldots, d\}$ such that $v_i < 0$:
$
m_i + v_i \leq (x \cdot a)_i \leq M_i, \; \forall x \in [m, M],
$
and for $i \in \{1, \ldots, d\}$ such that $v_i > 0$:
$
m_i \leq (x \cdot a)_i \leq M_i + v_i, \; \forall x \in [m, M].
$
Thus 
$||\newS - \newI||_1 \leq ||M-m||_1 + ||v||_1.$

Suppose that for at least one component $i \in A$, we
have two states $x$ and $y$ such that $x_i + v_i \in [0,C_i]$ and $y_i
+ v_i \not \in [0, C_i]$.  
(Otherwise $||\newS - \newI||_1 \leq ||M-m||_1$.)
For this component: $(\newS_i - \newI_i \leq M_i-m_i+ |v_i|-1$.
Indeed, suppose that $v_i > 0$ (the case $v_i < 0$ is symmetrical).
Then $M_i + v_i - 1 \geq C_i$ and $m_i \leq (z \cdot a)_i \leq C_i, \forall z
\in [m, M]$. 
Thus $\newS_i - \newI_i \leq C_i - m_i \leq M_i - 1 + v_i -m_i$, so 
we have:
$$||\newS - \newI||_1 \leq 
||M-m||_1 + ||v||_1 - 1.$$
\end{proof}

\section{Partition of the state space}
\label{sec:partition}

\subsection{Zone partition}

In this section we focus on a single event $a$. 
The state space is {\it partitioned into zones}, $Z^1,\cdots, Z^K$,
where $K$ is assumed to be small compared to the size of the state space.
The action $a$ may have a different effect $a^k$ on each zone $Z^k$.
Formally, the action $`\cdot`$ by $a$ is defined for all $x \in \Space$ as :
$x \cdot a = x \cdot a^k$ where $k$ is such that $x \in Z^k$.
To compute the envelopes
of the action of $a$ over an arbitrary interval
$[m,M]$,
one can use the partition into zones.

\begin{align}
%  \label{eq:part}
\nonumber
  [m,M]\edot a & = \Lb [m,M]\cdot a \Rb \\
\nonumber
&= \Lb  \left(\bigcup_{k=1}^K  [m,M]\cap Z^k \right) \cdot a \Rb \\
\nonumber
&= \Lb \bigcup_{k=1}^K  \left( [m,M]\cap Z^k  \cdot a\right)  \Rb \\
\nonumber
&= \Lb \bigcup_{k=1}^K  \left( [m,M]\cap Z^k  \cdot a^k
  \right)\Rb \\ 
&= \Lb \bigcup_{k=1}^K  \Lb  [m,M]\cap Z^k  \cdot a^k
  \Rb \Rb \label{eq:tight} \\ 
&\subset  \Lb \bigcup_{k=1}^K  \Lb  [m,M]\cap Z^k \Rb  \edot a^k
  \Rb.  \label{eq:computable}
\end{align}

The construction of these envelopes is illustrated in Figure \ref{fig-ce}.
The first and second equalities are direct consequence of  the definition of the partition;
the third  one is by definition of the action on a set;
the fourth  one simply uses the fact that $\Lb \Lb A \Rb \cup \Lb B \Rb
\Rb =  \Lb A\cup B \Rb$.

The last inclusion \eqref{eq:computable} does not hold with equality in general.
It uses the fact that for any set $A$, $\Lb A\cdot a \Rb \subset \Lb A\Rb \edot a$.
It gives us an easy way to compute an over-approximation of $\edot$.
Moreover, in some cases the inclusion can be replaced by an equality, such that
the over-approximation is exact (or {\it tight}). 
We give some simple cases where this equality holds in Section \ref{sec:tightness}.

\begin{figure}
  \centering
  \includegraphics{Envelope2.fig}  
  \caption{The construction of envelopes on a simple example.
The initial interval $[m,M]$ intersects the two zones $Z^1$ and $Z^2$.
We compute the next interval $[m',M'] \bydef [m,M] \edot a$ by computing
$[m^1,M^1] = \Lb [m,M] \cap Z^1 \cdot a \Rb$, 
$[m^2,M^2] = \Lb m,M \cap Z^2 \cdot a \Rb$,  
and finally $[m',M'] = \Lb [m^1,M^1] \cup [m^2,M^2] \Rb $ 
correspond to \eqref{eq:tight}, but in this example the equality
holds also in the inclusion \eqref{eq:computable}:
$[m',M'] = \Lb \Lb [m,M] \cap Z^1 \Rb \edot a \cup \Lb [m,M] \cap Z^2 \Rb \edot a \Rb $. 
 } 
  \label{fig-ce}
\end{figure}

The computation of an interval containing $[m,M]\edot a$ using \eqref{eq:computable} 
reduces to two problems:
\begin{itemize}
\item[$(P_1)$] The computation of  $\Lb  [m,M]\cap Z^k \Rb$. This  can be done 
  by integer linear programming (see next section).
\item[$(P_2)$] The computation of  $\Lb  [m,M]\cap Z^k \Rb  \edot a^k$.
This is easy when $a^k$ is an ASHE (using Algorithm \ref{al:ashes}) or when
$a^k$ is monotone (see Equation \eqref{eq:mono}).
\end{itemize}

In Section \ref{sec:lp}, we will give methods to solve $(P_1)$ 
by computing $\Lb  [m,M]\cap Z^k \Rb$ or an over-approximation.

\subsection{Tightness} \label{sec:tightness}

\begin{proposition} \label{prop:nonblocking}
  Let $U$ be a subset of $\Space$ and $a$ an ASHE.
  If $a$ is {\em non-blocking} on $U$ (\ie $\forall x \in U, B(x)=\emptyset$),
  then $ \Lb U \cdot a \Rb = \Lb U \Rb \edot a$.
\end{proposition}

\begin{proof}
The inclusion $\Lb U \cdot a \Rb \subseteq \Lb U \Rb \edot a$ is simple, since
by definition $\Lb U \Rb \edot a = \Lb \Lb U \Rb \cdot a \Rb$ and $U \subseteq \Lb U \Rb$.

Let us prove now the other inclusion. Let $[m,M]=\Lb U \Rb$.
Since we assumed $a$ to be non-blocking on $U$, \ie $\bigcup_{x \in U}B(x)=\emptyset$,
we have for all $x$ in $U$, $x \cdot a = (x+v)\vee 0 \wedge C$.
Using Lemma \ref{lem:subset}, we also have
$B(m)\cup B(M) = \bigcup_{x \in U}B(x) =  \emptyset$, which gives 
$[m,M] \edot a = [(m+v)\vee 0 \wedge C,(M+v)\vee 0 \wedge C]$
(see Algorithm~\ref{al:ashes}, in which
$B(m)\cup B(M) = \emptyset$ implies that $X = \emptyset$).

We can now prove that $ \Lb U \Rb \edot a \subseteq \Lb U \cdot a \Rb $.
Let $[m',M']=\Lb U \Rb \edot a$ and $[q,Q]=\Lb U \cdot a \Rb$.
To prove the inclusion we just need to show that $m' \ge q$ and $M' \le Q$.

We showed that $m'=(m+v)\vee 0 \wedge C$.
For each component $i \in \{1,\dots,d\}$, there exists $x \in U$ such that $x_i=m_i$.
Let $x'=(x+v) \vee 0 \wedge C$, we have $x' \in U \cdot a$, thus $x' \ge q$, and
$m_i' = m_i+v_i \vee 0 \wedge C_i = x_i+v_i\vee 0 \wedge C_i = x_i' \ge q_i$.
Finally, $m' \ge q$, and similarly $M' \le Q$.
\end{proof}

\begin{corollary}
  Let $a$ be a piecewise ASHE over a partition $Z^1, \dots ,Z^K$ such that
  each ASHE $a^k$ is non-blocking on its zone $Z^k$. Then the inclusion \eqref{eq:computable}
  can be replaced by an equality.
\end{corollary}

\begin{remark}
  It is easy to see that any ASHE $a$ can be partitioned into zones such that
  $a$ is non-blocking on each zone.
  Although this is not useful for computing the envelope of a simple ASHE
  (since we already know how to compute it), it can however be used to obtain 
  tight envelopes on piecewise ASHEs.
\end{remark}

\begin{proposition} \label{prop:unidimensional}
  Let $U$ be a subset of $\Space$, $a$ an ASHE and $v$ its direction vector.
  If $a$ is {\em unidimensional}, (\ie $Act=\{i \mid v_i \neq 0\}$ is a singleton),
  and if the projection of $U$ on any component $i$: 
  $U_i \bydef \{ x_i \mid x \in U \}$ is an interval,
  then $ \Lb U \cdot a \Rb = \Lb U \Rb \edot a$.
\end{proposition}

\begin{proof}
  The inclusion $\Lb U \cdot a \Rb \subseteq \Lb U \Rb \edot a$ 
  is always true, and was already proved. 
  We keep the same notation as in the previous proof: 
  $[m,M]=\Lb U \Rb$, $[m',M']=\Lb U \Rb \edot a$
  and $[q,Q]=\Lb U \cdot a \Rb$.
  Let $i$ be the component where $v_i \neq 0$. For all $j \neq i$, we have $m_j'=m_j$,
  $M_j'=M_j$, and for all $x \in U$, $(x \cdot a)_j=x_j$, thus $m_j'=q_j$ 
  and $M_j'=Q_j$ clearly.

  On component $i$, we need to consider several cases (we will show 
  that $m_i' \ge q_i$ and $M_i' \le Q_i$):
  \begin{itemize}
    \item If $i \notin B(m) \cup B(M)$, then the transition is non-blocking
      on $U$, and the previous Proposition applies.
    \item If $i \in B(m) \cap B(M)$, then $[m',M']=[m,M]$ and $U \cdot a = U$, thus
      $[m',M']=[q,Q]$.
    \item If $i \in B(m) \setminus B(M)$, then $m_i<-v_i<M_i$.
      In that case, we have $m_i'=0$ and $M_i'=(M_i+v_i)\vee(-v_i-1)$.  
      By assumption, $U_i=\{ x_i \mid  x \in U \}$ is an interval, 
      and thus $U_i=[m_i,M_i]$.
      Since $m_i<-v_i<M_i$, there exists $x \in U$ such that $x_i = -v_i$.
      We then have $(x \cdot a)_i = x_i+v_i = 0$, and thus $q_i=0=m_i'$.

      We now prove that $Q_i \ge M'_i = (M_i+v_i)\vee(-v_i-1)$.
      Since $U_i=[m_i,M_i]$, there exists $y\in U$ such that $y_i=M_i$, 
      thus $Q_i \ge (y\cdot a)_i=y_i+v_i=M_i+v_i$.
      Also, since $m_i \le -v_i-1 < M_i$, there exists a $z \in U$ 
      such that $z_i=-v_i-1$. We then have $z_i+v_i=-1<0$, and thus $i \in B(z)$ 
      and $(z \cdot a)_i=z_i=-v_i-1$, and finally $Q_i \ge (z \cdot a)_i \ge -v_i-1$.

    \item If $i \in B(M) \setminus B(m)$ a similar reasoning shows the same result.

  \end{itemize}

  We finally obtain the reverse inclusion 
  $\Lb U \Rb \edot a \subseteq \Lb U \cdot a \Rb $ . 

\end{proof}

\section{Polytopic Zones and Linear Programming}
\label{sec:lp}

In this section we consider a  class of piecewise events  for
which the computation of envelopes can  be done in a reasonable
amount of time and memory space.
We will focus here on one event $a$
assumed to be defined piecewise over polytopic zones. 

Again, the state space is a product $\Space = [0,C_1]  \times \dots \times [0,C_d]$, 
and is here partitioned into zones defined by linear constraints.

Let us consider $H$ hyperplanes of $\R^d$ in general position and assume
they do not intersect $S$ (by translating them by a small $\varepsilon$ if needed). 
These hyperplanes define a partition of $\R^d$
in $K$ convex polytopes $(P^1,...,P^K)$, and these polytopes induce 
a partition of the state space $S$ in $K$ zones $Z^i = \Space \cap P^i$.
Without loss of generality, we consider only the nonempty zones and assume
we have $K$ of them. 

Finally, over each zone $Z^k$, the event $a^k$ is 
assumed to be easily computable (e.g. an ASHE).

There  exists a relation between the total number of hyperplanes $H$
needed to define the zones and the maximal number of zones that can be defined  using them.
The number $K$ of zones is smaller than 
$1+ H + \binom{H}{2} + \dots +  \binom{H}{d}$.
This is because the state space $\Space$ may intersect all zones
(bounded or not) formed by $H$ hyperplanes in a general position.
The number of such zones is given by the value of the characteristic
polynomial of the hyperplanes at point -1 (see \cite{Zaslavsky}).

\begin{figure}[t]
  \centering
  \includegraphics{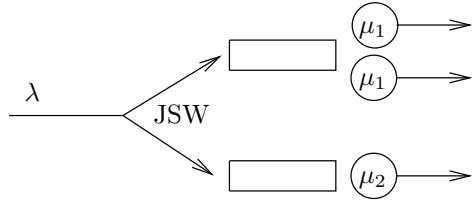}
  \caption{Example of a queueing event that  can be defined with ASHEs on polytopic zones.
  The first queue has two identical servers with rate $\mu_1$ while the second has
  only one server with rate $\mu_2$.}
  \label{jsw}
\end{figure}

\begin{figure}[t]
  \centering
  \includegraphics[width=.75\textwidth]{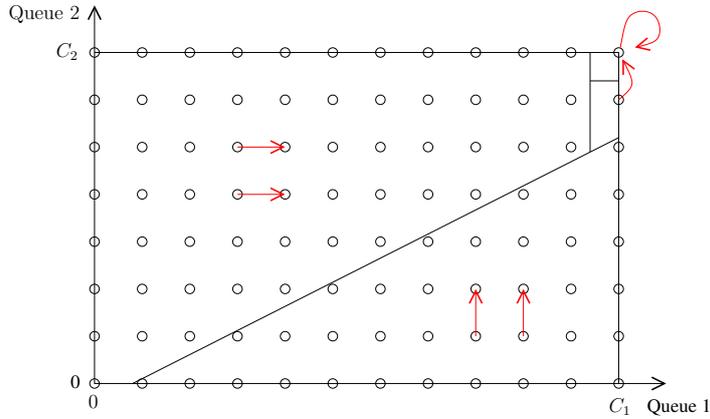}
  \caption{The zones for the JSW event  when $(2 \mu_1 > \mu_2)$.}
  \label{jswZones}
\end{figure}

\begin{example}
\label{ex:jsw}
A queueing example is displayed in Figure \ref{jsw}. We consider  a Join
the Shortest Waiting time (JSW) event, that is  an arrival 
that  allocates the new customer to the queue with the shortest
expected waiting time (after arrival). 
The zones for the JSW event are displayed in Figure
\ref{jswZones}. If $(x_2+1) / \mu_2  >  (x_1+1) / (2\mu_1) + \varepsilon $  
(with $\varepsilon$ very small) then the new customer is 
sent to queue 1 as long as the queue is not full, in which case it is sent to queue 2.
If $(x_2+1)/\mu_2  \le (x_1+1) / (2\mu_1) + \varepsilon $, it is sent to queue 2 as long
as the queue is not full, in which case it is sent to queue 1. 
If both queues are full, the client is rejected.
Here 3 hyperplanes divide the two-dimensional state space into 4
zones that are polytopes:
$ (x_2+1)/\mu_2 =  (x_1+1)/(2\mu_1) + \varepsilon$, $x_1 =  C_1 -1 +\varepsilon$, 
$x_2  = C_2 -1+\varepsilon$. 

Note that in this case, the partition makes the event a non-blocking ASHE in each zone,
so that Equation \eqref{eq:computable} holds with equality 
(by Proposition \ref{prop:nonblocking}).
However, the two zones in the upper right corner could be merged in one single
zone with a blocking ASHE, and the equality would still hold 
(see Proposition \ref{prop:unidimensional}).
\end{example}

\subsection{Zone intersection with Linear Programming}

\begin{figure}[ht]
  \centering
  \includegraphics[width=.6\textwidth]{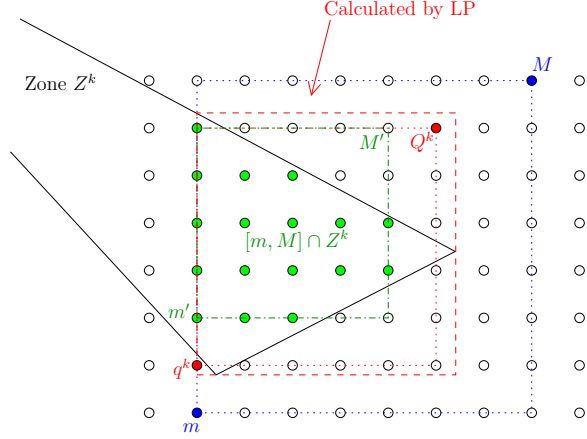}
  \caption{Consider a zone $Z^k$ and an interval $[m,M]$.
The Figure displays the intervals $[m,M]$, $[m',M'] \bydef \Lb  [m,M] \cap Z^k  \Rb$
and the larger interval $[q^k,Q^k]$ constructed using \eqref{eq:lp}.} 
  \label{intersect}
\end{figure}

Since we assumed the event $a$ to be easily computable on each zone $Z^k$,
the only problem left to solve is $(P_1)$ the computation of $[m',M'] = \Lb [m,M] \cap Z^k \Rb$.
When the zones $Z^k$ are induced by polytopes $P^k$ ($Z^k = \Space \cap P^k$),
this computation amounts to solving $2d$ integer linear programs, 
one for each component $m'_i$ and $M'_i$ :

\begin{equation}
  \forall i \in \{1,\dots,d\},~ \quad
  \begin{cases}
    m'_i & = \min \{ x_i \mid x \in \Z^d \cap P^k \cap [m,M] \}, \\
    M'_i & = \max \{ x_i \mid x \in \Z^d \cap P^k \cap [m,M] \}.
  \end{cases}
\end{equation}

Unfortunately, integer linear programs are well known to be NP-hard in general.
These linear programs can be solved in simple cases that we will not detail in this paper.
A general approach to overcome this difficulty is to relax the integrity condition and 
to solve the corresponding rational linear programs, by computing $q^k$ and $Q^k$ 
in $\Z^d$ defined by

\begin{equation}\label{eq:lp}
  \forall i \in \{1,\dots,d\},~ \quad
  \begin{cases}
    q^k_i & \bydef \lceil \min \{ x_i \mid x \in \Q^d \cap P^k \cap [m,M]_\Q \} \rceil, \\
    Q^k_i & \bydef \lfloor \max \{ x_i \mid x \in \Q^d \cap P^k \cap [m,M]_\Q \} \rfloor, 
  \end{cases}
\end{equation}
where $[m,M]_\Q$ denotes the rational interval $\{ x \in \Q^d \mid m \le x \le M\}$,
that  contains $[m,M]$.

 It is easy to see that the discrete interval $[q^k,Q^k]$ 
contains $\Lb [m,M] \cap Z^k \Rb$. 
We can thus compute an over-approximation of $\edot$ for $a$ :
 
 \begin{align*} \label{eq:continuous}
   [m,M]\edot a 
   & \subset \Lb \bigcup_{k=1}^K  \Lb  [m,M]\cap Z^k \Rb  \edot a^k \Rb \\
   &\subset \Lb \bigcup_{k=1}^K [q^k,Q^k] \edot a^k
   \Rb. 
 \end{align*}

Figure \ref{intersect} gives an illustration on a simple example.

\subsection{Complexity and data structure}

In order to deal with the set of linear programs in an efficient way,
they can be represented as a binary acyclic graph.
Each node corresponds to an additional  linear constraint. The left child
corresponds to the half plane where the inequality is negative and the right child 
to the one where the inequality is positive.

When an intersection can be reached through several paths, the nodes
are merged.
 
\begin{figure}[h]
  \centering
  \includegraphics[width=.9\textwidth]{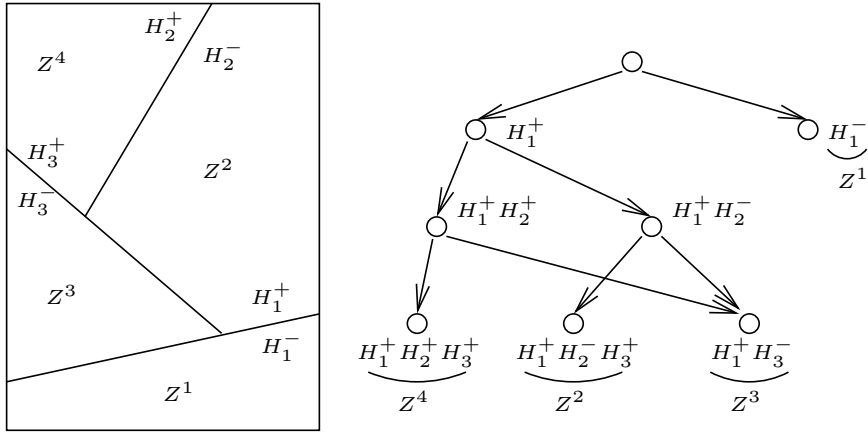}
  \caption{A partition and the corresponding graph structure used to sweep the zones. 
    Here, the space is partitioned into 4 polytopic zones, $Z_1, Z_2,Z_3$ and $Z_4$, 
    defined by three hyperplanes $H_1,H_2$ and $H_3$. 
    The corresponding half-spaces are denoted $H_i^+$ and $H_i^-$, for all $i$.
    Hence, for instance, $Z^2 = H_1^+ \cap H_2^- \cap H_3^+$.  }
  \label{fig:data}
\end{figure}

The leaves of the graph are all the polytopes and  the number of
nodes in the tree is less than  twice the number of leaves.
This construction is illustrated in Figure \ref{fig:data}.\

This representation has a better amortized complexity to compute
$[m,M] \cdot \nolinebreak a$ than an exhaustive resolution for each polytopic zone since
the intersection of $[m,M]$ with a zone can be found empty early in
the tree (this can even happen at the root of the tree if $[m,M]$ does
not intersect one half plane.

Sorting the constraints (from the root of the tree to the leaves)
to get the best possible expected cost of computing $[m,M]\cdot a$  can be
done for a given distribution of $m$ and $M$ (for example uniform over $\Space\times \Space$)
by using dynamic programming.  
This is not in  the scope of this paper.

\begin{algorithm}[htb]
\KwData{Partition of the state space into $Z^1,\ldots, Z^K$ arranged in a graph
structure $G$ using linear constraints;}
\KwResult{ interval $[m',M'] = \Lb
  \underset{P^k \cap [m,M]_\Q \neq \emptyset}{\bigcup_{k=1}^K}  
  [q^k,Q^k] \edot a^k \Rb. $ }
Bound ($V,m,M,a$):=\\
\Begin{
Visited($V$):= True\;
\If {$V \cap [m,M]_\Q \not= \emptyset$}
{\If {$V = P^k$}{
\For{$j = 1$  {\em  \textbf{to}} $d$}
   {$q^k_j$ := $\lceil$ LP-Solve($\min \{x_j \mid x\in V\cap [m,M]_\Q \}$)$ \rceil$\;
$Q^k_j$ := $\lfloor$ LP-Solve($\max \{x_j \mid x\in V\cap [m,M]_\Q \} $)$ \rfloor$\;
}
$[m',M']:= \Lb [m',M'] \cup [q^k,Q^k]\edot a^k \Rb$ ; 
                 }
\Else{\If {Visited($V$.left) = False}{Bound($V$.left,$m,M,a$)}
      \If {Visited($V$.right) = False}{Bound($V$.right,$m,M,a$)}
     }
}
}
\caption{Recursive procedure computing the bounds on the envelopes
  using linear programming}
\label{al:bound}
\end{algorithm}

The Algorithm \ref{al:bound} defines the procedure  Bound($V,m,M,a$)  
such that, when $V$ is the graph $G$, 
$\mbox{ Bound }(G,m,M,a)  =  \Lb \bigcup_{k=1}^K  [q^k,Q^k]  \edot a^k \Rb $ 
which is an over-approximation of $[m,M]\edot a$.
It uses the procedure LP-Solve (not detailed) that solves linear
programs over rational numbers in polynomial time.
It also contains the  test $V \cap [m,M]_\Q \not= \emptyset$ 
that  can be done much faster (in linear time, see the next section)
and helps to cut the cost: The LP-solve procedure is not used when the
intersection is empty.

By solving linear programs with the interior point method \cite{Karmarkar84},
the complexity of each main call to Algorithm \ref{al:bound} is $O(d^{4.5}K)$.

\subsection{Fast bounds using Minkowski sums}

The previous  cost to compute over-approximations of envelopes 
by solving linear programs may not be sustainable for chains with a large dimension $d$ and 
when  the number of calls to {\it Bound} is large.

In this section we will derive a way to compute looser bounds on the 
envelopes, with a
complexity $O(dHK)$. This technique is based on the
computation of Minkowski  sums of all $K$ zones with  intervals $[m, M]$
(see~\cite{Oks96}).

\begin{figure}[htb]
  \centering
  \includegraphics{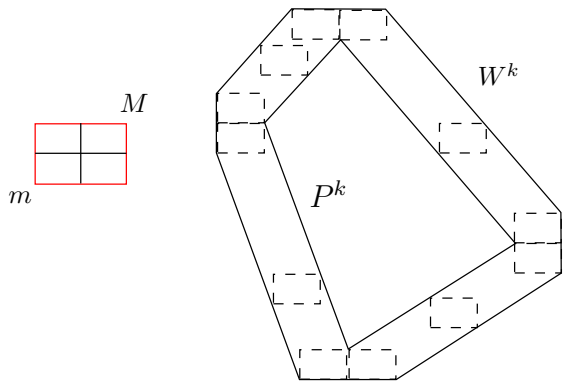}
  \caption{The Minkowski sum of a polytope $P^k$ and the
    interval $[-s/2,s/2]$, $s$ being the size of the interval $[m,M]$.}
  \label{fig-minkowski}
\end{figure}

Each polytope $P^k$ can be expressed as the intersection of $h_k$
hyperplanes: $P^k = \{ x \in \R^d \mid x A^k \le b^k \}$, where $A^k$
is a $d \times h_k$ matrix and $b^k$ a row vector of size $h_k$.
Each constraints $\{ A^k_{(\cdot,j)}x \le b^k_j \}$ correspond to one
of the hyperplane defining $P^k$.

Testing if an interval  $[m,M]_\Q$ of size $s \bydef M - m$  
intersects a polytope $P^k$ can be done by testing 
if the center of the interval ($m+s/2$) 
belongs to an extended zone $W^k$ that  is the
{\it Minkowski sum} of $P^k$ and the interval $[-s/2,s/2]$:
$W^k \bydef P^k \oplus [-s/2,s/2]$ 
(the Minkowski sum $A\oplus B$ of two sets $A$ and $B$ is defined by
$A\oplus B = \{ a+b | a\in A, b\in B\} $).

Figure \ref{fig-minkowski} shows an illustration of the Minkowski sum in a simple case.

It is direct to see that
\[ [m,M]_\Q \cap P^k \not= \emptyset \Leftrightarrow m+s/2 \in W^k.\]

and that the set $W^k$ is a polytopic zone defined by $h_k + 2d$ hyperplanes:
\begin{equation*}
W^k = \{ x \in \R^d \mid  xA^k \leq b^k + (s/2) |A^k| \}
 \cap \{ x \in \R^d \mid x  \ge \ell^k - s/2 ,x \le L^k + s/2 \}, 
\end{equation*}
where $|A^k| \bydef (|A^k_{i,j}|)_{1 \le i \le d, 1 \le j \le h_k}$  is the component-wise 
absolute value of matrix $A^k$, and $\ell^k$ and $L^k$ are defined in $\Q^d$ by

\begin{equation}\label{eq:rational_lp}
  \forall i \in \{1,\dots,d\},~ \quad
  \begin{cases}
    \ell^k_i & \bydef \min \{ x_i \mid x \in \Q^d \cap P^k \}, \\
    L^k_i & \bydef \max \{ x_i \mid x \in \Q^d \cap P^k \}.
  \end{cases}
\end{equation}

Note that $\ell^k$ and $L^k$ do not depend on $[m,M]$ and thus can be pre-computed
with linear programming before starting the simulation.

Testing if the  center of $[m,M]$ belongs to $W^k$ simply  amounts to
testing if the center satisfies all inequalities defining $W^k$. This
is done in time $O(dh_k)$, and $H$ being the total number of hyperplanes used
to define the polytopes, we have $h_k \le H$.
Therefore, the test $[m,M] \cap P^k \not= \emptyset$ in Algorithm \ref{al:bound} 
can be done in time $O(dH)$ for each polytope $P^k$.

The previous interval $[q^k,Q^k]$ computed with linear programs (solvable in $O(d^4)$) 
can be replaced by a looser over-approximation of $\Lb Z^k \cap [m,M] \Rb$ computable 
in time $O(2d)$.

Indeed, since $P^k \subset [\ell^k,L^k]_\Q$, we have 
\[ \Lb Z^k \cap [m,M] \Rb \subset [ \lceil \ell^k \rceil , \lfloor L^k \rfloor ] \cap [m,M] 
= [m \vee \lceil \ell^k \rceil , M \wedge \lfloor L^k \rfloor ],  \]
where $\lceil \ell^k \rceil$ (resp.~$\lfloor L^k \rfloor$) 
is the component-wise ceiling (resp.~floor) of $\ell^k$ (resp.~$L^k$) and is pre-computed.
The intervals $[m \vee \lceil \ell^k \rceil , M \wedge \lfloor L^k \rfloor]$ 
can replace the intervals $[q^k,Q^k]$ in Algorithm \ref{al:bound},
and are easier to compute. The complexity of Algorithm \ref{al:bound} would then be $O(dHK)$.
However, these intervals provide looser bounds on the envelopes
and might thus increase the coupling time of the bounding processes, 
at worst to infinity.
As we pointed out earlier, this tradeoff between
coupling time and complexity of computations needs to be adjusted 
depending on the Markov chain to simulate.

\section{Queueing Network Examples}
\label{sec-ex}

In this section, we illustrate the usefulness and the expressive power of 
piecewise ASHEs by showing examples of queueing networks that  can be described
using  Markov automatons with piecewise ASHEs over polytopic zones.

The rule of thumb is that many classical (and some non-classical) queueing
networks are such chains so that  sampling their  stationary distribution is made possible
using ASHES over polytopic zones, while the classical PSA would be intractable in practice.

\subsection{Jackson Network with finite capacities}
\label{ss:ex_jk}

We consider open queueing systems similar to 
Jackson networks except that queues have finite capacities.
The network is formed by $d$ queues numbered from 1 to $d$: 
Each queue $i$ is a M/M/1/$C_i$ queue with finite capacity $C_i$ and a FIFO server.
Entering customers arrive in each server $i$ according to a Poisson process
of parameter $\lambda_i$, and each server is assumed to have exponential service time $\mu_i$. 
After being served at queue $i$, a customer is directed to queue $j$ with probability $P_{i,j}$
or leaves the network with probability $P_{i,0} \bydef 1-\sum_{j=1}^d P_{i,j}$.
We will consider two types of blocking policy:
\begin{itemize}
  \item Client Loss (CL):
    If a client served by queue $i$ is sent to a full queue, it is lost.
  \item Restart Service (RS):
    If a client served by queue $i$ is sent to a full queue, it restarts its service at queue $i$.
\end{itemize}
We assume that new clients arriving in the network from outside are lost 
when sent to a full queue.
The state of the network is entirely described by the number of customers in each queue,
and the corresponding Markov chain can be generated with the following Markov 
automaton $\A=(\Space,A,D,\cdot)$ where
$\Space=[0,C_1]\times\dots\times[0,C_d]$, $A=\{a_{i,j} | 0 \le i,j \le d\}$,    
$D(a_{i,j})=\lambda_{a_{i,j}}/\sum_{a \in A} \lambda_a  $, where $\lambda_a$ the frequency
rate of each event $a$ is defined as:
\begin{equation*}
  \begin{cases}
    \lambda_{a_{0,0}}:=0 & \\
    \lambda_{a_{0,j}}:=\lambda_j & \forall j \in [1,\dots,d] \\
    \lambda_{a{i,j}}:=\mu_i P_{i,j} & \forall i,j \in [1,\dots,d]
  \end{cases}
\end{equation*}
and where the $a_{i,j}$ are ASHEs with direction vector 
$v_{a_{i,j}}=e_j-e_i$ for all $i,j \in \{0,\dots,d\}$ 
(with the convention $e_0=(0,\dots,0)$), and with a blocking relation
$\mathcal{R}_{a_{i,j}}$ depending on the blocking policy used:
$\mathcal{R}^{CL}_{a_{i,j}}=\{(i,j)\}$ for the Client Loss policy and
$\mathcal{R}^{RS}_{a_{i,j}}=\{(i,j),(j,i)\}$ for the Service Restart policy.
Note that when $i=0$ or $j=0$, the choice of $\mathcal{R}$ has no incidence. 
The blocking policy may also differ for each couple $i,j$.

It is easy to check that the events of this Markov automaton are all monotone
for the product order on $\Space=[0,C_1]\times[0,C_d]$.
Therefore for this type of networks, the coupling times for PSA and EPSA are equal.

\subsection{Examples of non-monotone ASHEs}

In this section, we  consider  other types of routing events that are not monotone
nor anti-monotone and can be described with ASHEs.
These events can easily be added to any Jackson networks with finite capacity,
and they can easily be combined with each other to give other types of ASHEs. 

Envelopes for these  events were already computed in \cite{BGV08} using
ad-hoc methods, while the ASHE framework now embraces them all.
The numerical results for these events are taken from \cite{BGV08}.

\paragraph{Fork and Join}
%\label{sec:forkjoin}

A common type of non-monotone event used in queueing network is fork and join.  
A fork queue is such that, at service completion, 
several customers are sent simultaneously  to several disjoint output queues.
A join is a queue with several input buffers. The server is only active when
no buffer is empty. A service completion removes one customer in each 
input buffer and sends one (merged) customer to an output buffer.

Fork events were already introduced in Example \ref{ex:fork} and shown to be ASHEs.
A join event with two input buffers $i$ and $j$ and output buffer $k$ is described by an ASHE
with direction vector $v=e_k-e_i-e_j$ and blocking relation depending on the blocking policy:
$\mathcal{R}^{CL}=\{(i,k),(j,k),(i,j),(j,i)\}$ or 
$\mathcal{R}^{RS}=\{(i,k),(j,k),(i,j),(j,i),(k,i),(k,j)\}$. 

Join events are not monotone for the natural product order: 
Consider for instance the join event $a$ with
$i=1,j=2$ and $k=3$ in a 3-server network, and consider the two states 
$x=(0,1,0)$ and $y=(1,1,0)$. Then $x < y$ while
$x\cdot a=(0,1,0)$ and $y \cdot a=(0,0,1)$ are incomparable.
Similarly, it is easy to show that fork is in general not monotone.

\paragraph{Negative customers}

We consider 
the additional events 
corresponding to  negative customers. 
After queue $i$ finishes a service, a negative customer is sent to queue $j$ 
and kills one customer in that queue, if there is any.
This event can be represented by an ASHE of direction vector 
$v=-e_i-e_j$ and blocking relation $\mathcal{R}=\{(i,j)\}$.
Such events are not monotone for the natural product order.

\begin{figure}[h]
\centering
  \includegraphics[width=0.7\textwidth]{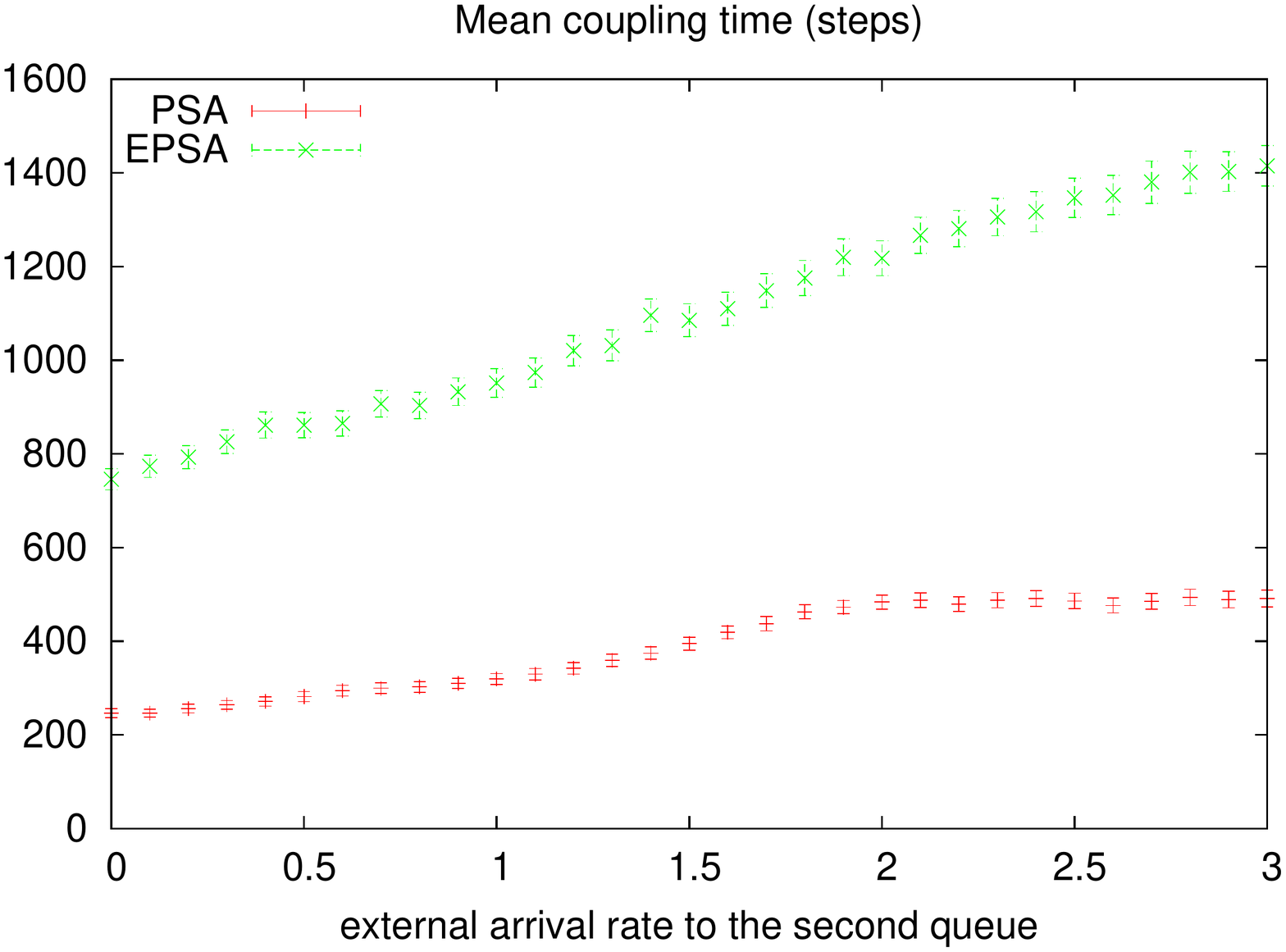}
  \caption{Mean coupling times of PSA and EPSA algorithms for the network in
    Figure \ref{fig:net1} as a function of~$\lambda_2$ with 95 \% confidence intervals.}
  \label{fig:neg}
\end{figure}

\begin{figure}[h]
  \centering
  \includegraphics[width=.7\textwidth]{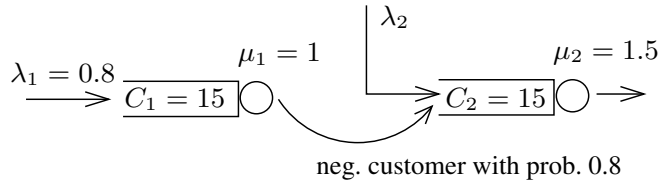}
  \caption{A network with negative customers.}
  \label{fig:net1}
\end{figure}

We ran simulations for the small network displayed in Figure
\ref{fig:net1}. The average coupling time of PSA and EPSA
are reported in Figure \ref{fig:neg} %below
(in the following all data are given with 95 \% confidence intervals).

In the example, the coupling time of EPSA is around 3 times larger than PSA. 
Since the number of states is $15^2 = 225$, this means that EPSA is 
about $225/6 = 40$  times faster than PSA in this case.

%\begin{figure}[htb]
%  \centering
% \includegraphics[width=.70\textwidth]{negEnvVSEx.pdf}
%\caption{Mean coupling times of PSA and EPSA algorithms for the network in
%  Figure \ref{fig:net1} as a function of~$\lambda_2$ with 95 \% confidence intervals.}
%\label{fig:neg}
%\end{figure}

\paragraph{Batch routing}
\label{sec:batch}

Here, we consider batch routing: After the end of service in queue $i$, a batch of $K$ customers 
leaves queue $i$ and creates $L$ new customers in queue $j$. In general, $K$ and $L$ do not need
to be equal: for instance $10$ products that
are assembled into one ({\em e.g.} put into a box), and then sent to next phase of a 
production line can be modeled by taking $K=10$ and $L=1$. 
This event can be represented by an ASHE of direction vector 
$v = -K e_i + L e_j$ and blocking relation depending on the blocking policy: 
$\mathcal{R}^{CL}=\{(i,i),(i,j)\}$ or
$\mathcal{R}^{RS}=\{(i,i), (i,j), (j,i), (j,j)\}$. As before, the exogenous arrivals 
(resp. services) can be modeled by setting $i=0$ (resp. $j=0$).
We allow partial arrivals to queue $j$ (in the case of $\mathcal{R}^{CL}$), 
while the service must be complete (\ie at least $K$ customers must be present in queue $i$).
However, partial service can also be modeled by piecewise ASHEs, using similar ideas as for 
multiple servers in Section \ref{ss:ex_pw}.

\begin{figure}[h]
  \centering
  \includegraphics[width=0.68\textwidth]{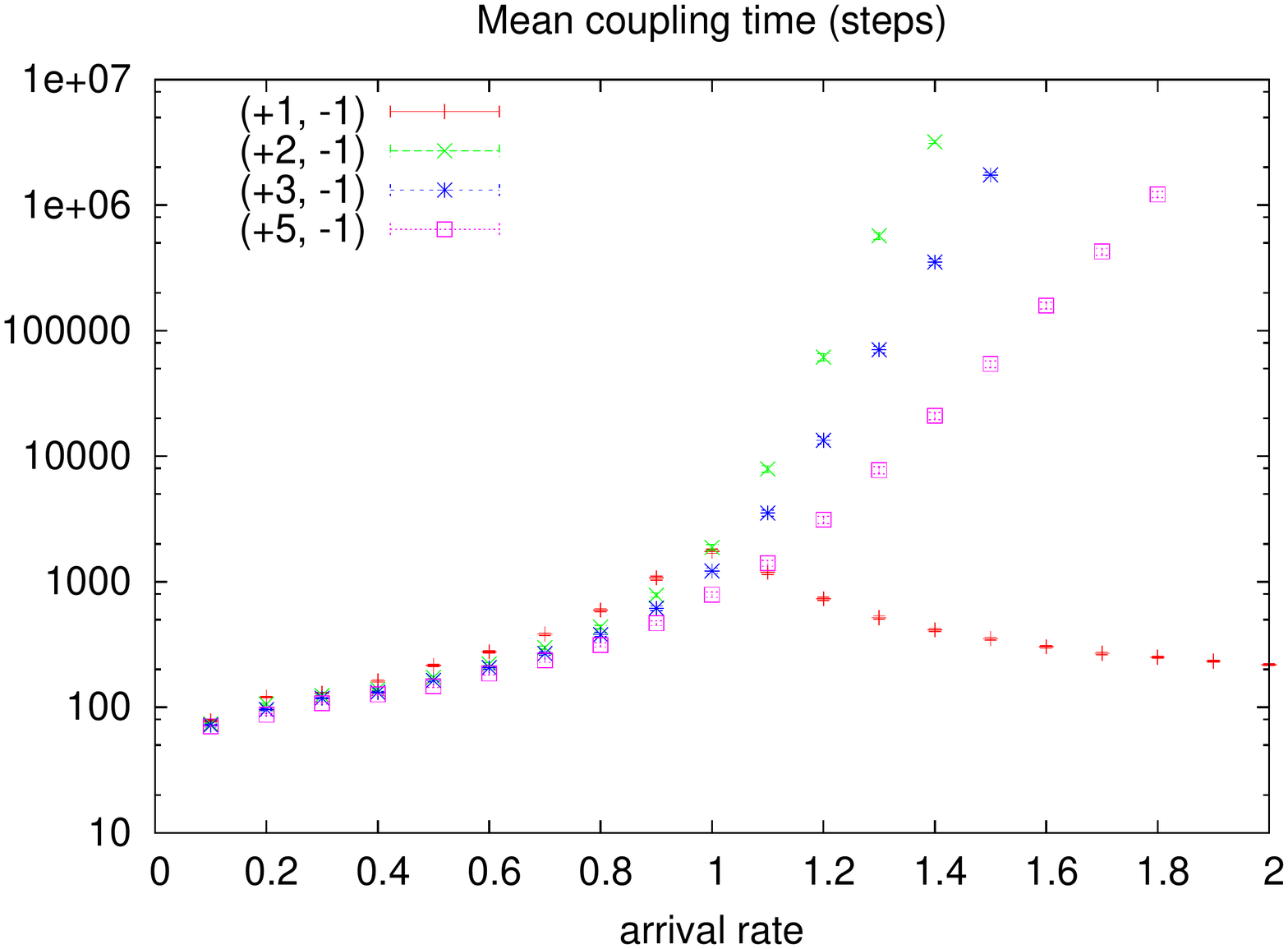}
  \caption{Mean coupling times for EPSA in a $(+k,-1)$ queue with batch arrivals of
    size $k = 1, 2,3$ and $5$ respectively, as a function of the arrival rate~$\lambda$.}
  \label{fig:batch1}
\end{figure}

Figure \ref{fig:batch1} reports the coupling time of EPSA  for a
single queue with batch arrivals of size $K = 1,2,3$ or $5$ with rate $\lambda$, 
departures of size $1$ and rate $1$, and blocking policy $\mathcal{R}=\{(1,1)\}$.
Unlike what happens for negative customers and fork-join nodes,
the coupling time of EPSA increases exponentially fast with the load
of the queue (the scale being logarithmic).
Such increase is due to the fact that envelopes can only
couple after they both hit 0.
The time for the upper envelope starting in $C$ to hit $0$ when the
load $\rho$ is larger than 1 becomes exponentially large (larger than
$e^{\rho C}$).

One way to prevent this is to use the Split algorithm \cite[Algorithm 3]{BGV08},
that is an hybrid between PSA and EPSA: It starts by computing bounding
intervals as EPSA until the size of the interval
becomes smaller than the size of the largest batch.
When this happens, the bounding interval is then split into all the single
trajectories contained in it, and these normal trajectories are then computed as in PSA.
The advantage of the Split algorithm is that these normal trajectories as not as
numerous as the $|\Space|$ trajectories needed for the classical PSA,
and it is guaranteed to terminate when PSA does unlike EPSA.

\begin{figure}[h]
  \centering
  \includegraphics[width=0.68\textwidth]{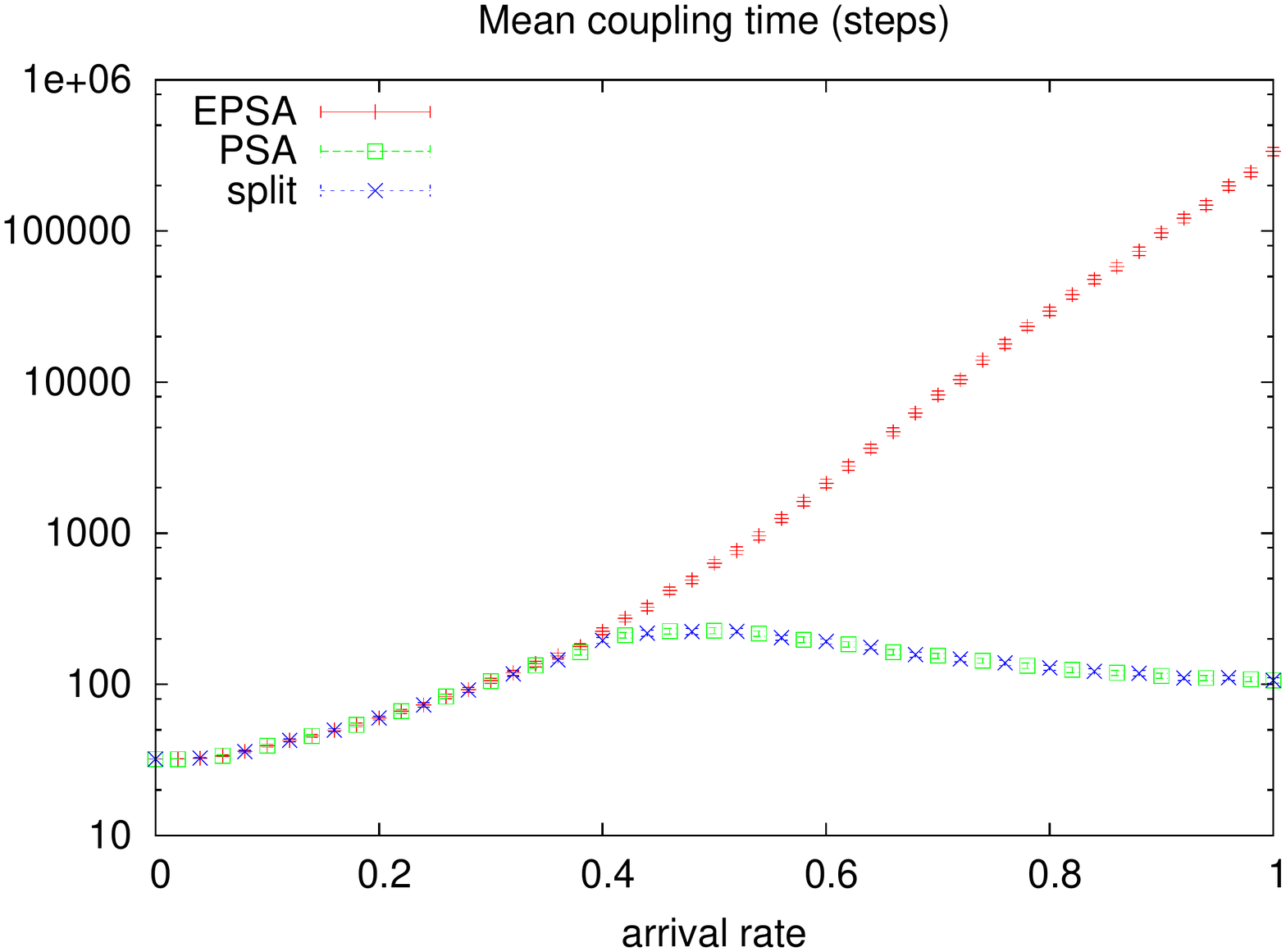}
  \caption{Mean coupling times for PSA, EPSA and Split for a $(+2, +3, -1)$ queue.}
  \label{fig:batch32}
\end{figure}

Figure \ref{fig:batch32} reports the coupling times of EPSA, PSA
and Split for a queue with capacity $20$, arrival batches $+2$ and $+3$
with probabilities $0.49$ and $0.51$, and  departures of size one (with rate $1$).
In this figure, the coupling time of PSA and Split are
similar while EPSA's coupling time grows very fast when $\lambda/\mu$ is larger than one.
In this case Split has a clear advantage over the other
two algorithms:
It couples exponentially faster than EPSA and deals with two
envelopes instead of $C$ trajectories for PSA.
The efficiency of Split can be partially explained by the 
choice of the splitting point: 
as long as the size of the bounding interval is 
larger than $K$ (the size of a batch), 
the batch arrival event is nonexpansive. 
This is a direct consequence of Proposition~\ref{prop:rej}.

\subsection{Examples of Piecewise ASHEs}
\label{ss:ex_pw}

We now consider  even more complex queueing events that will prove to be 
piecewise ASHEs over polytopic zones.

\paragraph{Multiple Servers}

\begin{figure}[t]
   \centering
     \subfloat[Transition rate matrix]{
       \includegraphics[width=0.85\textwidth]{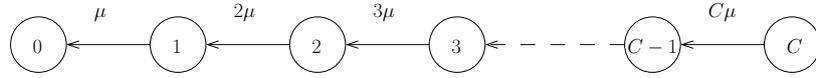}
     }  \\
     \subfloat[Corresponding Markov automaton]{
       \includegraphics[width=0.85\textwidth]{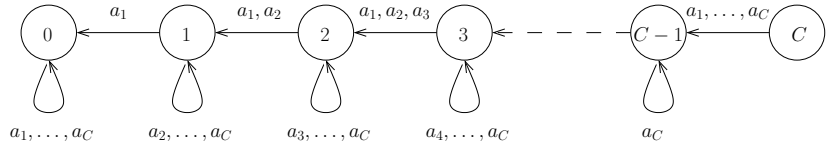}
     }
   \caption{Representation of the transition rate matrix (a) of the multiserver M/M/$C$/$C$
     queue, and its corresponding description as a Markov automaton (b) using $C$ piecewise ASHEs
     $a_1,\dots,a_C$, each being drawn with rate $\mu$.
   }
   \label{fig:multiple_servers}
 \end{figure}

Let us assume now that queue $i$ is a M/M/$C$/$C$ queue with finite capacity $C$ and $C$
identical servers with serving rate $\mu$, and that customers are sent to queue 
$j$ after being served. 
Such type of service can be described using $C$ piecewise ASHEs $a_1,\dots,a_C$
as illustrated in Figure~\ref{fig:multiple_servers}.

The state space of queue $i$ is $[0,C]$ and the transition rate matrix $Q$
associated to this event is such that $Q(x,x-1)=x\mu$, the other coordinates
being null.
We add $C$ events $a_1,\dots,a_C$ with each $a_k$ defined as the following piecewise ASHE: 
on zone $[0,k-1]$, $a_k$ is the identity ($v=0$, $\mathcal{R}=\emptyset$)
and on $[k,C]$, $a_k$ is the ASHE with direction vector ($v=e_j-e_i$)
and blocking relation $\mathcal{R}=\emptyset$. 
Finally, we set $\lambda_{a_k}=\mu$, and take $D(a_k)=\lambda_{a_k}/{\sum_{a \in A}\lambda_a}$.

This construction can be easily adapted to any queue with $n < C$ servers, which  can be represented with $n$ piecewise ASHEs.
It can also be easily combined with any other type of events  discussed previously.

~

\paragraph{Join the Shortest Waiting Time (JSW) and other routing policies}

JSW is described in Example \ref{ex:jsw} and is shown 
to be a piecewise ASHE over polytopic zones.

More generally, %any type of deterministic routing can be described as a piece-wise ASHE. 
consider the routing event $r_i$ where a client, after being
served at queue $i$, is sent to queue $j=R(x)$ for some function $R: \Space \rightarrow [0,d]$ 
(if $j=0$ the client leaves the system),
and the client is either lost (CL) or restarts its service (RS) if the destination queue is full.
Such event can be described as a piecewise ASHE over at most $d$ zones
$Z^k=R^{-1}(k)$ for $k=1,\dots,d$ with the event on $Z^k$ being an
ASHE with direction vector $e_{k} - e_i$ and blocking relation $\mathcal{R}^{CL}\{(i,k)\}$
or $\mathcal{R}^{RS}=\{(i,k),(k,i)\}$ depending on the blocking policy.
Provided the zones are polytopic, it is then easy to compute 
over-approximation of envelopes for this event.
For some non-polytopic zones $Z$, the interval $\Lb[m,M]\cap Z\Rb$ can
still be easily computed with had-oc methods.
For instance, if we consider the routing by index
$R(x)=\argmin_{i \in \{0,\dots,d\}} f_i(x_i)$ for some 
strictly increasing real functions $f_i$ (we assume the $f_i$
functions are chosen such that the $\argmin$ is unique for any $x \in \Space$),
then the corresponding zone is of the form 
$Z^k=R^{-1}(k)=\{x \mid \forall j \in \{1,\dots,d\},~ f_{k}(x_k) \le f_{j}(x_j) \}$,
and one can show that $[m',M']=\Lb[m,M]\cap Z^k\Rb$ is
given by $m'_k=m_k$, $M'_k=f_k^{-1}(\min_{i \in \{0,\dots,d\}} f_i(M_i)) $,
and $\forall i \neq k$, $m'_i=f_i^{-1}(f_k(m_k)) \vee m_i$ and
$M'_i=M_i$.

Of course, routing policies can be combined with
all the previous types of event (forks and joins, negative customers, batches, multiple servers)
to give more types of piecewise ASHEs.

\subsection{Application to a network}

Here is an example of a network where events are ASHEs or piecewise ASHEs over polytopic zones.
The following model permits us to compare the routing policies JSW to the Random (Rand) routing.

In this model, we consider two parallel queues $1,2$ with same capacity $C_1=C_2=20$ and 
one server with service rate $\mu_1,\mu_2$, respectively. We introduce $\mu=\mu_1+\mu_2$
and $a=\mu_1/\mu$ such that $\mu_1=a\mu$ and $\mu_2=(1-a)\mu$.
Clients arrive in the system at a rate $\lambda=1$ and go in one of the two queues,
according to the chosen routing policies.

\begin{figure}[ht]
   \centering
   \includegraphics[width=0.9\textwidth]{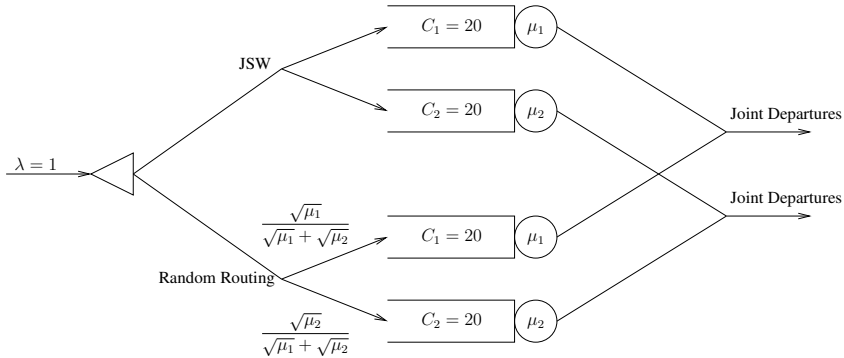}
   \caption{Network to compare the JSW and the Random routing policies}
   \label{fig:fork_jsw}
 \end{figure}

To compare the two policies, the 2-queue system is duplicated to give a 4-queue-network,
with two 2-queue subsystems with coupled arrivals and departures (see Figure \ref{fig:fork_jsw}).
Incoming clients are duplicated via a fork and sent to each subsystem. 
If one of the subsystem rejects its copy of the client, the other must reject it as well.
This is to ensure that both systems always receive the same quantity of clients.

The first system then routes its clients using the JSW policy, while the second system sends its
clients randomly to queue $1$ with probability  $\frac{\sqrt{\mu_1}}{\sqrt{\mu_1}+\sqrt{\mu_2}}$
and to queue $2$ with remaining probability. If the packet arrives in a full queue, it is rejected
(and the client is rejected by the other subsystem as well).
The departures are also coupled: with rate $\mu_1$ [resp.~$\mu_2$], 
a packet leaves queue $1$ [resp.~$2$] (if the queue is not already empty), 
\textit{simultaneously for each subsystem}. 
When the queue is empty in one subsystem, the corresponding queue in the other subsystem
can still serve its clients.

\begin{figure}[t]
  \centering
  \includegraphics[width=.48\textwidth]{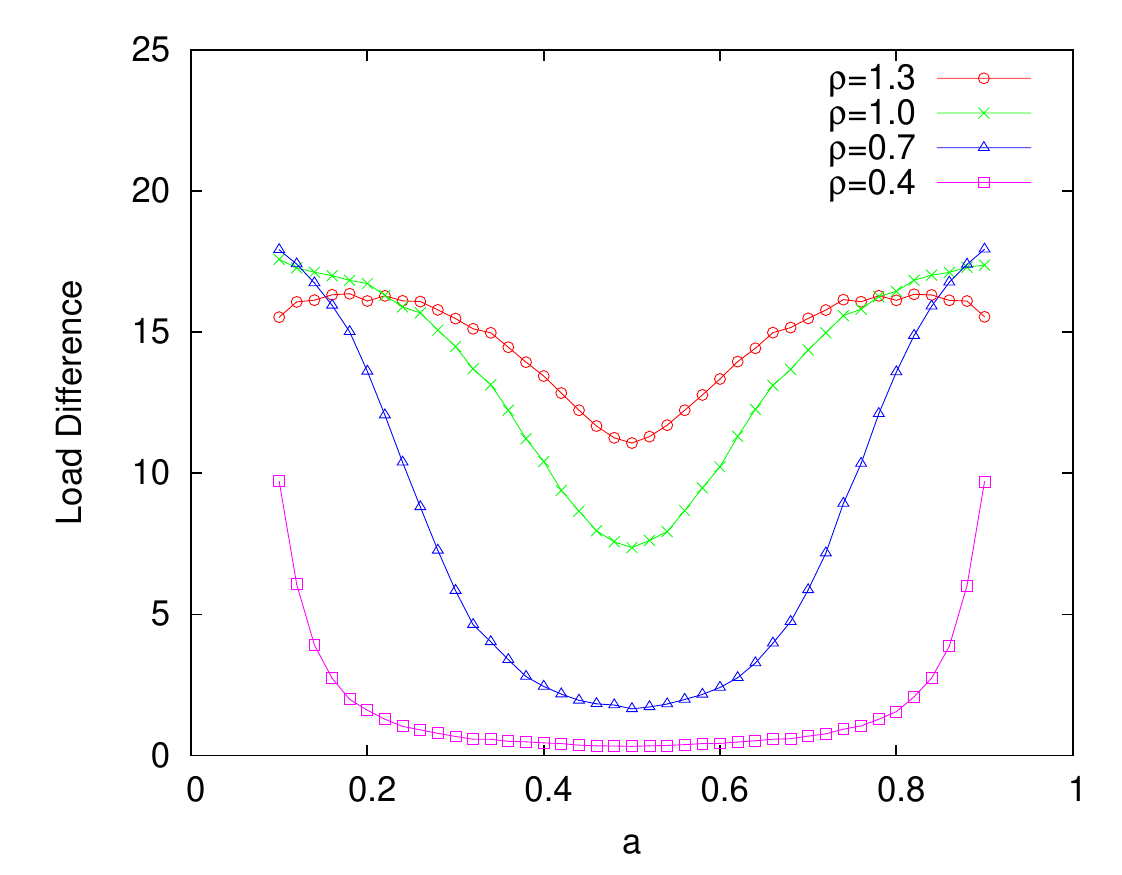}
  \includegraphics[width=.49\textwidth]{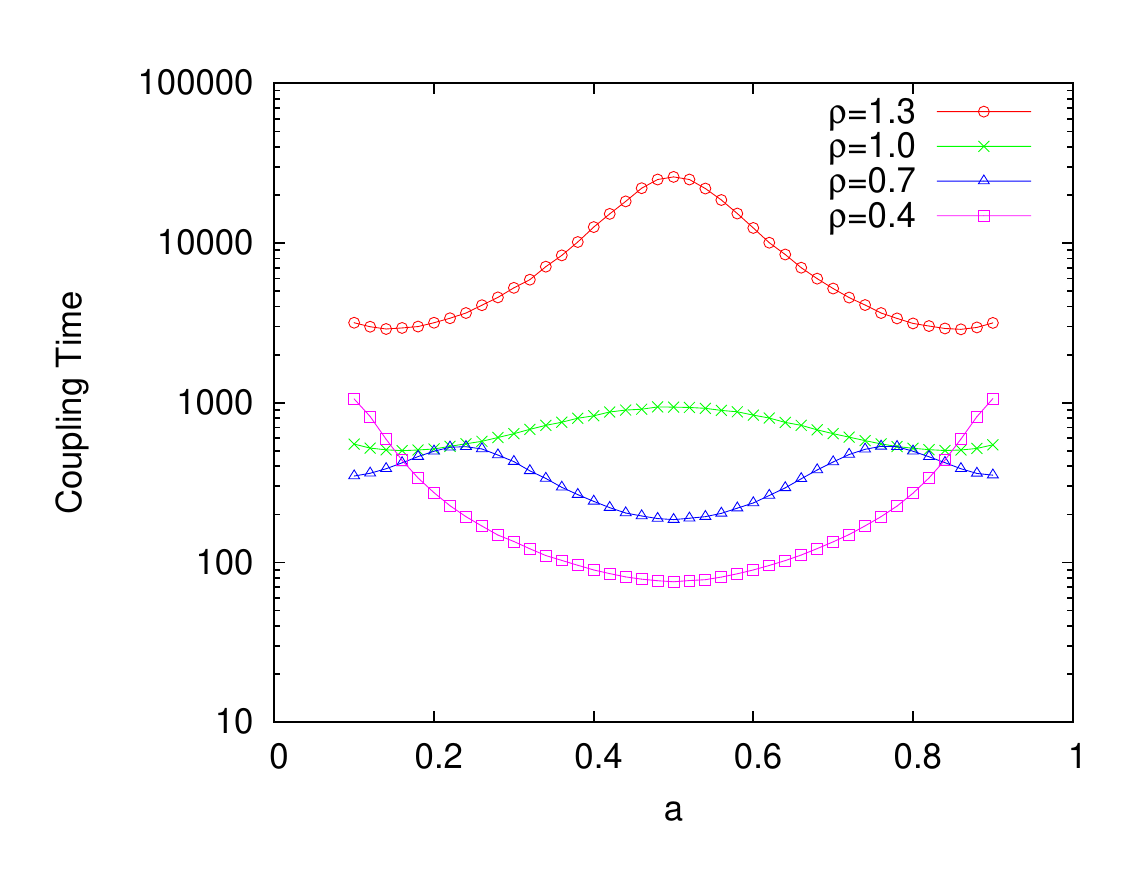}
  \caption{
On the left: load difference between the two subsystems 
(Difference between the average number of clients in the random routing servers 
and the JSW servers) when $a$ varies and for several value of the network load~$\rho$.
On the right: corresponding coupling time (number of iteration). Note that the scale
is logarithmic in the number of iterations. For each point of the figure, 10000 simulations
were run.}
  \label{fig:model2} 
\end{figure}

Figure \ref{fig:model2} shows the results of the  simulation of that network 
using the envelope technique. We plotted the average difference of clients in the queue 
between the two subsystems, when the ratio $a$ varies between $0.1$ and $0.9$,
and for several values of the network load $\rho=\lambda/\mu$.

As we can see, the JSW routing strategy performs better than the random 
routing in all cases, and particularly when the serving rates are not 
close from each other or when the workload is high.

The coupling time is much lower for low load. 
In that case the dominant events are non-blocking services. 
If the load is high, the dominant events are arrivals that  are blocking. 
Further, the coupling can only be achieved  by hitting the lower boundary 
of the state space (by emptying the system), so the coupling time grows exponentially with the 
load.

\section{Conclusion}
\label{sec-conclusion}

Perfect sampling  can be a very efficient method to study the stationary 
behavior of a Markov chain. 
This efficiency relies on three conditions.
The first one concerns the small number of simulated trajectories needed for unbiased coupling. 
The second condition is the requirement that a  small time should be needed to compute those 
trajectories and the third condition is to have a small coupling time.
When these three conditions are satisfied, the perfect sampling technique 
is hard to beat to compute the stationary distributed with a given confidence interval.

In this paper we show that the first condition is easy to meet by using envelopes. 
We also show that the second condition can be met for piecewise space homogeneous chains by 
using  a special type of events (ASHEs) and by over approximating the envelopes.
The third condition is more chain dependent. We provide several examples where coupling using 
the envelope technique happens fast and others where the coupling time is very large. 
In those cases, the splitting technique can help reduce the coupling time without 
compromising  with the other two points.

\subsection*{Acknowledgments}

The authors wish to thank Steven Oudot 
for his useful remarks.

\bibliographystyle{abbrv}
\bibliography{space_homogen_MC}

\begin{thebibliography}{10}

\bibitem{BBT02}
F.~Baccelli, B.~B{\l}aszczyszyn, and F.~Tournois.
\newblock {Spatial averages of coverage characteristics in large CDMA
  networks}.
\newblock {\em Wirel. Netw.}, 8:569--586, 2002.

\bibitem{BGV08}
A.~Bu\v{s}i\'{c}, B.~Gaujal, and J.-M. Vincent.
\newblock Perfect simulation and non-monotone markovian systems.
\newblock In {\em Valuetools'08}, Athens, Grece, 2008.

\bibitem{cassandras}
C.~G. Cassandras and S.~Lafortune.
\newblock {\em Introduction to discrete event systems}.
\newblock Springer, 2nd edition, 2008.

\bibitem{Dopper}
J.~Dopper, B.~Gaujal, and J.-M. Vincent.
\newblock Bounds for the coupling time in queueing networks perfect simulation.
\newblock In {\em Numerical Solutions for Markov Chains (NSMC'06)}, pages
  117--136, Charleston, 2006.
\newblock The 2006 A.A. Markov Anniversary Meeting (MAM 2006).

\bibitem{KFPVV_psmsb_07}
J.-M. Fourneau, I.~Y. Kadi, N.~Pekergin, J.~Vienne, and J.-M. Vincent.
\newblock Perfect simulation and monotone stochastic bounds.
\newblock In {\em Valuetools'07}, Nantes, France, 2007.

\bibitem{GP_cfphmf_07}
B.~Gaujal and F.~Perronnin.
\newblock Coupling from the past in hybrid models for file sharing peer to peer
  systems.
\newblock In {\em Proceedings of the 10th International Conference {HSCC}'07},
  Pisa, Italy, 2007.

\bibitem{GaujalPerronninJDEDS}
B.~Gaujal, F.~Perronnin, and R.~Bertin.
\newblock Perfect simulation of a class of stochastic hybrid systems with an
  application to peer to peer systems.
\newblock {\em Journal of Discrete Event Dynamic Systems}, 2007.
\newblock Special Issue on Hybrid Systems.

\bibitem{HaNe98}
O.~H{\"a}ggstr{\"o}m and K.~Nelander.
\newblock Exact sampling from anti-monotone systems.
\newblock {\em Statist. Neerlandica}, 52(3):360--380, 1998.

\bibitem{Hu04}
M.~Huber.
\newblock Perfect sampling using bounding chains.
\newblock {\em Ann. Appl. Probab.}, 14(2):734--753, 2004.

\bibitem{Karmarkar84}
N.~Karmarkar.
\newblock A new polynomial time algorithm for linear programming.
\newblock {\em Combinatorica}, 4(4):373--395, 1984.

\bibitem{Kendall-1998d}
W.~S. Kendall.
\newblock {Perfect simulation for the area-interaction point process}.
\newblock In L.~Accardi and C.~C. Heyde, editors, {\em Probability Towards
  2000}, pages 218--234, New York, 1998. University of Warwick Department of
  Statistics, Springer-Verlag.

\bibitem{KendallMoeller-2000}
W.~S. Kendall and J.~M{\o}ller.
\newblock {Perfect simulation using dominating processes on ordered spaces,
  with application to locally stable point processes}.
\newblock {\em Advances in Applied Probability}, 32(3):844--865, 2000.

\bibitem{Oks96}
E.~Oks and M.~Sharir.
\newblock Minkowski sums of monotone and general simple polygons.
\newblock {\em Discrete and Computational Geometry}, 35(2):223--240, 1996.

\bibitem{propp96exact}
J.~G. Propp and D.~B. Wilson.
\newblock Exact sampling with coupled {M}arkov chains and applications to
  statistical mechanics.
\newblock {\em Random Structures and Algorithms}, 9(1-2):223--252, 1996.

\bibitem{V_psmsfr_05}
J.-M. Vincent.
\newblock Perfect simulation of monotone systems for rare event probability
  estimation.
\newblock In {\em Winter Simulation Conference}, Orlando, Dec. 2005.

\bibitem{dbwilson}
D.~B. Wilson.
\newblock Perfectly random sampling with {M}arkov chains.
\newblock Annotated bibliograpy.
  \url{http://dimacs.rutgers.edu/~dbwilson/exact.html/}.

\bibitem{Zaslavsky}
T.~Zaslavsky.
\newblock Facing up to arrangements: face-count formulas for partitions of
  space by hyperplanes.
\newblock {\em Memoirs of the American Mathematical Society (Providence, R.I.:
  American Mathematical Society)}, (154), 1975.

\end{thebibliography}

\appendix

\section{Proof of Theorem \ref{thm:ashes}}
\label{app:ashe}

Denote $[\newI, \newS] = [m, M]\edot a$. 
Consider a fixed component $j$. 
We have the following different cases: 
\begin{enumerate}
 \item $j \not\in \cup_{x \in [m,M]} B(x)$ (component $j$ is not blocked in any 
state $x$, $m \leq x \leq M$).  
By Lemma \ref{lem:subset}, this is equivalent to $j \not\in B(m) \cup B(M)$.
Then: 
$$ [m'_j,M'_j] = \left [(m_j + v_j) \lor 0 \land {C_j}, \;  (M_j
  + v_j) \lor 0 \land {C_j}\right ].$$ 
\item $j \in \cap_{x \in [m,M]} B(x)$  (component $j$ is blocked in
  all states $x$, $m \leq x \leq M$).  
It is easy to see that this occurs if and only if there is 
$i \in \CR(m) \cap \CR(M)$ such that $(i, j) \in {\cal R}$. 
Then $(x \cdot a)_j = x_j, \; m \leq x \leq M$ and:
 $$ [m'_j,M'_j] = \left [m_j, \;  M_j\right ].$$

\item  $j \in (\cup_{x \in [m,M]} B(x)) \backslash (\cap_{x \in [m,M]} B(x))$  
(component $j$ is blocked in some but not all states $x$, $m \leq x \leq M$). 
Then we have the following cases:

\begin{enumerate}
\item  There is  $i \not=j$ such that  $(i, j) \in {\cal R}$ and $i
  \in \CR(m) \cup \CR(M)$  
(component $j$ is blocked by a component $i$ different than $j$). 

If $v_j < 0$, then $(j,j) \not\in{\cal R}$ or $j \not\in \CR(M)$
(remark that  $(j,j) \in{\cal R}$ and $j \in \CR(M)$ would imply $j
\in \CR(m)\cap\CR(M)$ and thus 
$j \in \cap_{x \in [m,M]} B(x)$). We have:
$$ [m'_j,M'_j] = \left [(m_j + v_j)\lor {0}, \; M_j\right ].$$

If $v_j > 0$, then similarly $(j,j) \not\in{\cal R}$ or $j \not\in \CR(m)$ and:
$$ [m'_j,M'_j] = \left [m_j, \; (M_j + v_j) \land {C_j}\right ]. $$ 

\item  Otherwise (only $j$ is blocking $j$). Then $(j, j) \in {\cal
    R}$ and $j \in \CR(m) \cup \CR(M)$.    

If $v_j < 0$, then $j \in \CR(m)\backslash\CR(M)$ (\ie $m_j < - v_j
\leq M_j$) and:  
$$ [m'_j,M'_j] = \left [0, \;  (M_j+v_j) \lor (-v_j - 1) \right ].$$

Note that $-v_j-1 = \max \{ x_j \mid x \in [m,M],~ j \in B(x)\} $. 
%is the greatest value of $x_j$ for which the event is blocked. 

If $v_j > 0$, then $j \in \CR(M)\backslash\CR(m)$  (\ie $C_j - M_j <
v_j \leq C_j - m_j$) and:  
$$ [m'_j,M'_j] =\left [(m_j+v_j) \land (C_j-v_j+1), \; C_j\right ].$$ 

\end{enumerate}
\end{enumerate}

For a given component $j$, the complexity of computing $[m'_j, M'_j]$ amounts
to determining the corresponding case. However, this can be precomputed using 
$O(|{\cal R}|) = O(d^2)$ elementary operations. Thus the complexity of Algorithm \ref{al:ashes}
is $O(d^2)$.

\end{document}